\documentclass[useAMS,usenatbib]{mn2e}
\usepackage{graphicx}
\usepackage{journals}
\usepackage[hyperindex,pdftex]{hyperref}
 \usepackage{epstopdf}
 \usepackage{epsfig}
\usepackage{amsmath}
\usepackage{fixltx2e}
\usepackage{url}
\def\mysingleq#1{`#1'}
\def\mydoubleq#1{``#1''}
\newcommand{\hompc}{\,h\,{\rm Mpc}^{-1}}
\newcommand{\mpcoh}{\,h^{-1}\,{\rm Mpc}}
\newcommand{\gpcoh}{\,h^{-1}\,{\rm Gpc}}

\newcommand{\gpcohV}{\,h^{-3}\,{\rm Gpc^3}}
\newcommand{\hompcV}{\,h^{3}\,{\rm Mpc^{-3}}}
\newcommand{\simgt}%
{\,\hbox{\lower0.6ex\hbox{$\sim$}\llap{\raise0.6ex\hbox{$>$}}}\,}
\newcommand{\simlt}%
{\,\hbox{\lower0.6ex\hbox{$\sim$}\llap{\raise0.6ex\hbox{$<$}}}\,}

\title{Efficient Reconstruction of Linear Baryon Acoustic Oscillations in Galaxy Surveys.}
\author[A.Burden et al.]{
\parbox{\textwidth}{
A. Burden$^{1}$\thanks{E-mail: angela.burden@port.ac.uk}, 
W. J. Percival$^{1}$,
M. Manera$^{2}$,
Antonio J. Cuesta$^{3,4}$,
Mariana Vargas Magana$^{5}$,
Shirley Ho$^{5}$.\\
}
\vspace*{4pt} \\
$^{1}$Institute of Cosmology and Gravitation,  University of Portsmouth, Dennis Sciama building, Portsmouth, PO1 3FX\\
$^{2}$University College London, Gower Street, London, WC1E 6BT\\
$^{3}$Institut de Ci{\`e}ncies del Cosmos, Universitat de Barcelona,
IEEC-UB, Mart{\'\i} i Franqu{\`e}s 1, E-08028, Barcelona, Spain\\
$^{4}$Department of Physics, Yale University, 260 Whitney Ave, New
          Haven, CT 06520, USA  \\
$^{5}$Bruce and Astrid McWilliams Center for Cosmology, Department of
          Physics, Carnegie Mellon University,\\ 5000 Forbes Ave,
          Pittsburgh, PA 15213, USA }
          
\begin{document}

\date{}


\maketitle

\label{firstpage}

\begin{abstract}
Reconstructing an estimate of linear Baryon Acoustic Oscillations (BAO) from an evolved galaxy field has become a standard technique in recent analyses. By partially removing non-linear damping caused by bulk motions, the real-space BAO peak in the correlation function is sharpened, and oscillations in the power spectrum are visible to smaller scales. In turn these lead to stronger measurements of the BAO scale. Future surveys are being designed assuming that this improvement has been applied, and this technique is therefore of critical importance for future BAO measurements. A number of reconstruction techniques are available, but the most widely used is a simple algorithm that decorrelates large-scale and small-scale modes approximately removing the bulk-flow displacements by moving the overdensity field \citep{Eisenstein:2006nk,PhysRevD.79.063523}. We consider the practical implementation of this algorithm, looking at the efficiency of reconstruction as a function of the assumptions made for the bulk-flow scale, the shot noise level in a random catalogue used to quantify the mask and the method used to estimate the bulk-flow shifts. We also examine the efficiency of reconstruction against external factors including galaxy density, volume and edge effects, and consider their impact for future surveys. Throughout we make use of the mocks catalogues created for the Baryon Oscillation Spectroscopic Survey (BOSS) Date Release 11 samples covering $0.43<z<0.7$ (CMASS) and $0.15<z<0.43$ (LOWZ), to empirically test these changes.
\end{abstract}

\begin{keywords}
reconstruction, BAO, galaxy survey
\end{keywords}

\section{Introduction}
\label{sec:intro}
Many different scenarios have been proposed to explain the observed accelerated expansion rate of the Universe, based on perturbing either the matter-energy content of the Universe or the law of gravity away from the standard General Relativity + Cold Dark Matter picture. In order to differentiate between models, it is important to establish robust and accurate measurements of the expansion rate. The  Baryon Acoustic Oscillation (BAO) scale provides a standard ruler in the distribution of mass, and in turn galaxies, allowing a mechanism to make such measurements. The BAO feature arises from spherical imprints in the density field, remnants of pressure waves that travelled away from perturbations, through the tightly coupled photon, baryon plasma of the early Universe (e.g. \citealt{1999MNRAS.304..851M}). The scale of the pattern depends on the sound horizon at the baryon drag epoch - quantifying the distance propagated by the waves. For the fiducial concordance $\Lambda$CDM model that we adopt in this paper, the sound horizon $r_d=149.28$ \,Mpc (comoving),  which is close to the best fit value cited in \citet{2013arXiv1303.5076P}. 

In the correlation function of the matter density field, this effect leads to a peak at a scale corresponding to the sound horizon - where any perturbation is surrounded by a spherical shell of higher than average density. In the Fourier representation of the 2-point statistic, the power spectrum, the effect translates to a series of peaks and troughs as a function of scale. These patterns of density perturbations expand with the expansion of the Universe meaning the observed BAO scale in a galaxy distribution depends on the sound horizon projected at the redshifts of the galaxies, in the observed units of redshift and angle. Thus the BAO feature provides a mechanism to measure the combination of the sound horizon with the angular diameter distance $D_A(z)/r_d$ and Hubble parameter $H(z)r_d$ across and along the line-of-sight respectively \citep{2003ApJ...598..720S, 2003ApJ...594..665B, 2003PhRvD..68f3004H}.

For a sample of galaxy pairs with an isotropic distribution and clustering signal, the projection of the BAO peak in the monopole depends on $D_V(z)/r_d$, where
\begin{equation}
D_{\textrm{V}} \left(z\right) = \left[ cz \left(1 + z\right) ^2 D_A^2\left(z\right) H^{-1}\left(z\right)\right]^{1/3}.\label{eq:Dv}
\end{equation} 
Locations of peaks in the temperature-temperature Cosmic Microwave Background (CMB) power spectrum provide a similar measurement, where the projection depends on the angular diameter distance at the last scattering surface. A full fit to both CMB and galaxy survey data for a set of cosmological models provides further constraints on $r_d$, allowing accurate distance measurements to the survey redshifts. 

Recent measurements of the BAO scale in galaxy surveys have built up a distance ladder, mainly based on monopole measurements constraining $D_V(z)/r_d$ \citep{2010MNRAS.401.2148P, 2010ApJ...710.1444K, 2011MNRAS.416.3017B, 2011MNRAS.418.1707B, 2012MNRAS.427.2132P, 2012arXiv1203.6594A, 2013arXiv1312.4877A, 2014MNRAS.440.2222T}. At higher redshifts, measurements from the Ly-$\alpha$ forest have anchored this ladder at an epoch before Dark Energy \citep{2013JCAP...04..026S,2014arXiv1404.1801D,2014JCAP...05..027F}. The most recent data from the Baryon Oscillation Spectroscopic Survey (BOSS) are of sufficient quality that the measurement of $D_A(z)/r_d$ and $H(z)r_d$ from the monopole and quadrupole, provide enough extra information beyond monopole-only fits that the extra complication is warranted \citep{2013arXiv1312.4877A}.

In the power spectrum, the BAO signal continues to small scales (typical galaxy surveys contain a BAO signal to $k\simlt0.3\hompc$), extending from the linear into the non-linear regime, where the signal is degraded. This degradation increases in importance to low redshift, and results from increasing bulk motions of matter and non-linear structure formation \citep{0004-637X-664-2-660}. These processes move galaxies on average by approximately $10\mpcoh$ from their linear BAO positions resulting in a smearing of the acoustic feature in configuration space, which is equivalent to a damping of the BAO in the power spectrum \citep{1999MNRAS.304..851M,0004-637X-633-2-575,2005APh....24..334W}. This significantly reduces the precision of the BAO scale measurement.

This picture of the BAO signal is further complicated by Redshift Space Distortions (RSD; \citealt{1987MNRAS.227....1K}), which result from using the observed relative velocity of each galaxy to deduce the position. Peculiar velocities distort these positions from those due to cosmological expansion. RSDs induce a non-zero quadrupole moment in the measured density field. In the linear regime, they cause an increase in the amplitude of the power spectrum or correlation function monopole. On smaller, non-linear scales where velocities are incoherent with the large-scale structure, they generate an additional damping term. Thus the BAO damping is dependent on the angle to the line-of-sight for a redshift-space galaxy sample. The amplitude and signal-to-noise of the Fourier modes are also angle dependent.

As the signal degradation due to bulk flow is gravitationally induced, \citet{Eisenstein:2006nk} suggested it is possible to partially reverse this effect, utilising the galaxy map to estimate the potential that sources the motions between regions of a given scale. These motions can be used to mitigate the damping and, in effect, recover information about the linear overdensity.
The process is called reconstruction and has precursors dating back to \citet{1989ApJ...344L..53P}; see \citet{Eisenstein:2006nk} for a brief review of previous work. Most recent work to measure the BAO scale has used this simple algorithm for which a perturbation theory based analysis was presented by \citet{PhysRevD.79.063523} and extended to biased tracers in \citet{2009PhRvD..80l3501N}.

\begin{table*}
 \centering
 \begin{minipage}{150mm}
  \caption{Measurements from SDSS reconstructed galaxy surveys}
    \label{table:SDSSrecon} 
  \begin{tabular}{@{}llrrrrlrlr@{}}
  \hline
Reference  & Data Sample  & Pre Reconstruction Error & Post Reconstruction Error  \\
 \hline
 \citet{2013arXiv1312.4877A}    & DR11  CMASS       &   1.5\% &   0.9\%\\ %
 \citet{2014MNRAS.440.2222T}   & DR11 LOWZ          &   2.7\% &   1.9\%\\  %
 \citet{2014MNRAS.437.1109R}  & DR10 red sample &   2.7\%  & 2.0\% \\
 \citet{2014MNRAS.437.1109R}  & DR10 blue sample &   3.1\%  & 2.6\% \\
\citet{2013arXiv1312.4877A}    & DR10  CMASS  &  1.9\%  &  1.3\%\\%
\citet{2014MNRAS.440.2222T}   & DR10 LOWZ      &   2.6\% &   2.5\%\\  %
\citet{2012arXiv1203.6594A}       & DR9   &1.7\%  & 1.7\% \\
 \citet{2012MNRAS.427.2132P} & DR7  LRG  \footnote{The DR7 and DR9 constraints come from correlation function measurements whereas the DR10 and DR11 values quoted here are from the power spectrum measurements.}   &   3.5\%  &  1.9\% \\

\hline
\end{tabular}
\end{minipage}
\end{table*}

The reconstruction technique has been successfully applied to a number of galaxy samples selected from the Sloan Digital Sky Survey data (a list of results and references is provided in Table~\ref{table:SDSSrecon}) and also to the WiggleZ Dark Energy Survey \citep{2014MNRAS.441.3524K}. Reconstruction increased the precision of the measurements in all of the samples analysed, except for the DR9 CMASS sample \citep{2012arXiv1203.6594A} and the DR10 LOWZ sample \citep{2014MNRAS.440.2222T} where neither achieved a statistically significant improvement in the BAO scale measurement with reconstruction. Analysis with mock samples demonstrated that reconstruction is a stochastic process; reconstruction is less likely to reduce initially small errors. Both of these samples were \mydoubleq{lucky} data sets with a small pre reconstruction error. The pre reconstruction DR10 LOWZ error is smaller than the pre reconstruction DR11 LOWZ error although the sample covers a smaller volume and has a less contiguous area. 

Although the reconstruction algorithm suggested by \cite{Eisenstein:2006nk} is theoretically straightforward, it requires several assumptions. In this paper we empirically test these to establish the most efficient set of values to use. 
In Section \ref{sec:recon} we briefly review first order Lagrangian Perturbation Theory, and describe the practicalities of creating the reconstruction algorithm. In Section \ref{sec:mocks} we describe the simulations that we use to carry out our analysis. In Section \ref{sec:fitting} we describe the fitting procedure used to measure the BAO scale.
In Section \ref{sec:density} we look at how the survey density impacts the outcome, Section \ref{sec:edges} checks the effects of survey edges on results. Section \ref{sec:method} looks at various aspects of the method such as smoothing length, how many random data points are required, different ways of implementing the algorithm and removal of redshift space distortions, to see how these factors effect the performance of reconstruction. We present our conclusions in Section \ref{sec:conclusion}.

For efficiency we conduct our analysis in Fourier space using the power spectrum rather than the correlation function to measure the BAO. Previous analyses have shown the two methods to produce the same results \citep{2013arXiv1312.4877A, 2014MNRAS.440.2222T}. Throughout, we assume the cosmological model used to calculate the mocks,  $\Omega_m = 0.274$,  $h =0.7$, $ \Omega_b h^2=0.0224$,  $n_s = 0.95$ and $\sigma_8 = 0.8$.

\section{The Lagrangian reconstruction method}
\label{sec:recon}

The degradation of the BAO signal is expected to be dominated by bulk flows in the velocity field. While methods that alter the distribution of displacements while keeping the rank ordering the same can make the distribution look more like that of linear theory (e.g. \citealt{2012MNRAS.425.2443K}) they do not necessarily remove the small-scale damping. 
The method proposed by \citet{Eisenstein:2006nk} splits the density field in scale by moving densities according to displacements calculated from a smoothed field. In a Fourier framework, this reduces the damping of the oscillations due to bulk motions \citep{PhysRevD.79.063523}. In configuration-space one can see that densities on the smoothing scale are moved towards their ``linear'' positions by correcting the non-linear displacements at this scale.

We now review the algorithm, building up to the assumptions made when performing a practical implementation. The reconstruction method is based on estimating the displacement field from a smoothed version of the observed galaxy overdensity field. The galaxies, and points within a {\em random} catalogue that Poisson samples the 3D survey mask, are moved backwards based on this displacement field. We refer to these as the {\em displaced} and the {\em shifted} field respectively. The small-scale motions stay in the galaxy field, while the large scale clustering signal moves into the random catalogue. 2-point statistics are measured based on the difference between the galaxy and random fields. In Section~\ref{sec:displace_theory} we consider how the displacements are estimated, then in Section~\ref{sec:displace_practical} we discuss some of the practicalities of implementation.

\subsection{The observed galaxy displacement field in perturbation theory}  \label{sec:displace_theory}

It is natural to work in a Lagrangian frame work where the Eulerian position of a particle $\mathbf{x}$ can be described by the sum of its Lagrangian position $\mathbf{q}$ and some displacement vector $\mathbf{\Psi}$. 
\begin{equation}
\mathbf{x}\left(\mathbf{q},t\right) = \mathbf{q} + \mathbf{\Psi}\left(\mathbf{q},t\right).
\end{equation}
\citet{Eisenstein:2006nk} use the galaxy density field to estimate the Lagrangian displacements. To build up to this, we first review the first order Lagrangian Perturbation Theory (LPT) method of estimating the Lagrangian displacement field from a matter density field sampled at $\mathbf{x}$. 

Conservation of mass allows us to equate the total average density in Lagrangian coordinates with the sum of the Eulerian density,
\begin{equation}
\bar{\rho}d^{3}q = \rho\left(\mathbf{x},t\right)d^{3}x.
\end{equation}
where $\rho\left(\mathbf{x}\right)$ is the density of the matter at position $\mathbf{x}$ and $\bar{\rho}$ is the average density.
Thus the first order overdensity in Eulerian space can be related to the first order Lagrangian displacement vector by
\begin{equation}
\nabla_{\mathbf{q}}\cdot\mathbf{\Psi}_{\left(1\right)}\left(\mathbf{q},t\right) = -\delta_{\left(1\right)}\left(\mathbf{x},t\right),
\end{equation}
with the subscript $\left(1\right)$ as a reminder that they are both first order terms.
Assuming $\mathbf{\Psi}$ is an irrotational vector field \citep{1995A&A...296..575B}, it can be expressed in terms of a Lagrangian potential where
\begin{equation}\label{eq:potential}
\mathbf{\Psi}_{\left(1\right)}\left(\mathbf{q},t\right) = -\nabla_{q}\Phi\left(\mathbf{q},t\right),
\end{equation}
such that 
\begin{equation}
\label{eq:delPSI}
\nabla_{\mathbf{q}}\cdot\mathbf{\Psi}_{\left(1\right)}\left(\mathbf{q},t\right) = - \nabla_{q}^{2}\Phi\left(\mathbf{q},t\right)=-\delta_{\left(1\right)}\left(\mathbf{x},t\right).
\end{equation}
From these relations we can derive an expression for the first order displacement field in Fourier space that can be calculated directly from the Fourier transform of the overdensity field,
\begin{equation}
\mathbf{\Psi}_{\left(1\right)}\left(\mathbf{k}\right) = - \frac{i\mathbf{k}}{k^{2}}\delta_{\left(1\right)}\left(\mathbf{k}\right).
\end{equation}
This relation is the standard Zel'dovich approximation \citep{1970A&A.....5...84Z} and is the first order term in a Lagrangian perturbation theory expansion of the displacement field.

For a galaxy survey, we typically have to use the distribution of galaxies to estimate the matter field of the Universe, although this may change for future surveys with simultaneous weak-lensing and galaxy survey coverage. The current situation poses several problems:

The first is that galaxies are biased tracers of the matter. In this work we correct for this by assuming a local deterministic galaxy bias such that $\delta_g=b\delta$, where $b$, the galaxy bias, is the assumed ratio between the galaxy overdensity $\delta_g$ and matter overdensity $\delta$.

Secondly, 3D galaxy positions are inferred from their angular position on the sky combined with their redshift. Thus we have to assume a cosmological model for the distance-redshift relation before we can perform the reconstruction. However, the approximation of only performing reconstruction for a single fiducial model is expected to only weakly effect measurements: in \citet{2012MNRAS.427.2132P} they show that the distance scale measurement, $D_V/ r_s$, is robust to changes in the value of $\Omega_M$ used within a flat $\Lambda$CDM cosmology.

Thirdly, Redshift Space Distortions create a non-zero quadrupole moment with a sign dependent on whether they are in the linear/non-linear regime: linear RSD enhance the clustering signal along the line-of-sight, while incoherent non-linear peculiar velocities reduce it. The strength of linear redshift space distortions at a given redshift depends on the amplitude of the peculiar velocity field, and can be characterised by $f\sigma_8$, where $f= d \ln D\left(a \right) / d \ln a$, $D\left(a\right)$ is the growth function and $a$ is the scale factor.

To account for galaxy bias and RSDs, Eq.~(\ref{eq:delPSI}) can be modified following \citet{1994ApJ...421L...1N} and \citet{2012MNRAS.427.2132P} to
\begin{equation}
\label{eq:maineq}
\nabla \cdot \mathbf{\Psi} + \frac{f}{b} \nabla \cdot \left(\mathbf{\Psi\cdot \hat{r}}\right)\mathbf{\hat{r}} = -\frac{\delta_g}{b}.
\end{equation}
This is the first-order equation linking the displacement field to a sample of galaxies. An estimation of the potential can also be used to remove linear RSD from the galaxy distribution \citep{1987MNRAS.227....1K, 2004PhRvD..70h3007S, 2007ApJ...664..675E,2012MNRAS.427.2132P} by displacing the galaxies by an additional
\begin{equation} 
\label{eq:RSD}
\mathbf{\Psi}_{RSD} = -f\left(\mathbf{\Psi \cdot \hat{r}}\right)\mathbf{\hat{r}},
\end{equation}
where the $\mathbf{r}$ vector points along the radial direction of the survey. Note that this correction is not the same as removing the redshift space distortions in the Lagrangian displacement field as per Eq~(\ref{eq:maineq}) and removes the estimated RSD signal on a galaxy by galaxy basis.

\subsection{Practical implementation}  \label{sec:displace_practical}

Eq.~(\ref{eq:maineq}) can be solved either using finite difference techniques in configuration space or in Fourier space, where the vector operators have a simple form.
While \citet{2012MNRAS.427.2132P} used a finite difference method, we have considered both approaches and found them to match (see Section~\ref{sec:FD}). Our standard approach is to use Fourier based calculations on a Cartesian grid, which are computationally less expensive. 

The calculation of the smoothed overdensity from which the displacements are computed requires an estimate of the average galaxy density. This is commonly realised using a catalogue of random points Poisson sampled within the survey mask. As discussed above, a shifted random catalogue is also required which forms part of the reconstructed overdensity alongside the displaced galaxy catalogue. These catalogues should not be the same to avoid inducing spurious fluctuations between the derived potential and shifted fields. In order to minimise shot noise, the random catalogue should have a higher density than that of the galaxies: in this paper we use 100 times more randoms than galaxies for all tests, unless stated otherwise. To ensure that the randoms match the galaxy density as a function of redshift, $\bar{n}\left(z\right)$, we match the radial distributions after removing the redshift space distortions from the galaxies. We do this by assigning each random point a redshift picked at random from the galaxy catalogue post RSD removal. 

We carry out our tests on the SDSS III PTHalo mocks which are described in more detail in Section~\ref{sec:mocks}. The catalogues are set within boxes of length $3.5\gpcoh$. The overdensity and Fast Fourier Transforms (FFTs) are calculated on a $512^3$ grid. The size of the box is larger than the survey by at least $200\mpcoh$ on each side to ensure sufficient zero padding to avoid aliasing. A nearest grid-point assignment scheme is used to calculate the overdensity. We do not use any interpolation scheme to fill in regions within the box that are not covered by the survey as done in \citet{2012MNRAS.427.2132P}.

To ensure that the correct size regions source our Lagrangian displacement vectors, the density fields are convolved with a Gaussian filter, $S\left(k\right)=e^{-\left(k R\right)^2/2}$, where R is the smoothing length. This alleviates small scale non-linear motions, ensuring they do not contribute to the estimates of growth-related distortions. The convolution is carried out in Fourier space prior to calculating the overdensity. 

The smoothing can introduce spurious fluctuations in the overdensity outside of the survey volume. To account for this, we create a binary angular mask using the {\sc mangle} software \citep{2008MNRAS.387.1391S}. Imposing redshift cuts we create 3D mask used to cut the smoothed galaxy and random fields prior to calculating the overdensity. As the mask abruptly nulls the density of regions outside the survey, we find a slight deformation of the Lagrangian displacement field at the survey boundaries as well as the standard edge effects caused by loss of signal. We investigate these effects in Section~\ref{sec:edges}.

In order to use Eq.~(\ref{eq:maineq}), we need estimates for the values of bias $b$ and the growth factor $f$ for each galaxy catalogue to be analysed. While $f$ can be calculated for a fiducial cosmology, we must estimate $b$ empirically from the data itself. Thankfully, the measurements are insensitive to mild deviations as shown in Appendix B in \citet{2012arXiv1203.6594A}, although we would expect a loss in efficiency of the reconstruction algorithm for larger deviations.
\section{Mock catalogues}  \label{sec:mocks} 
\begin{figure}
  \includegraphics[width=\linewidth]{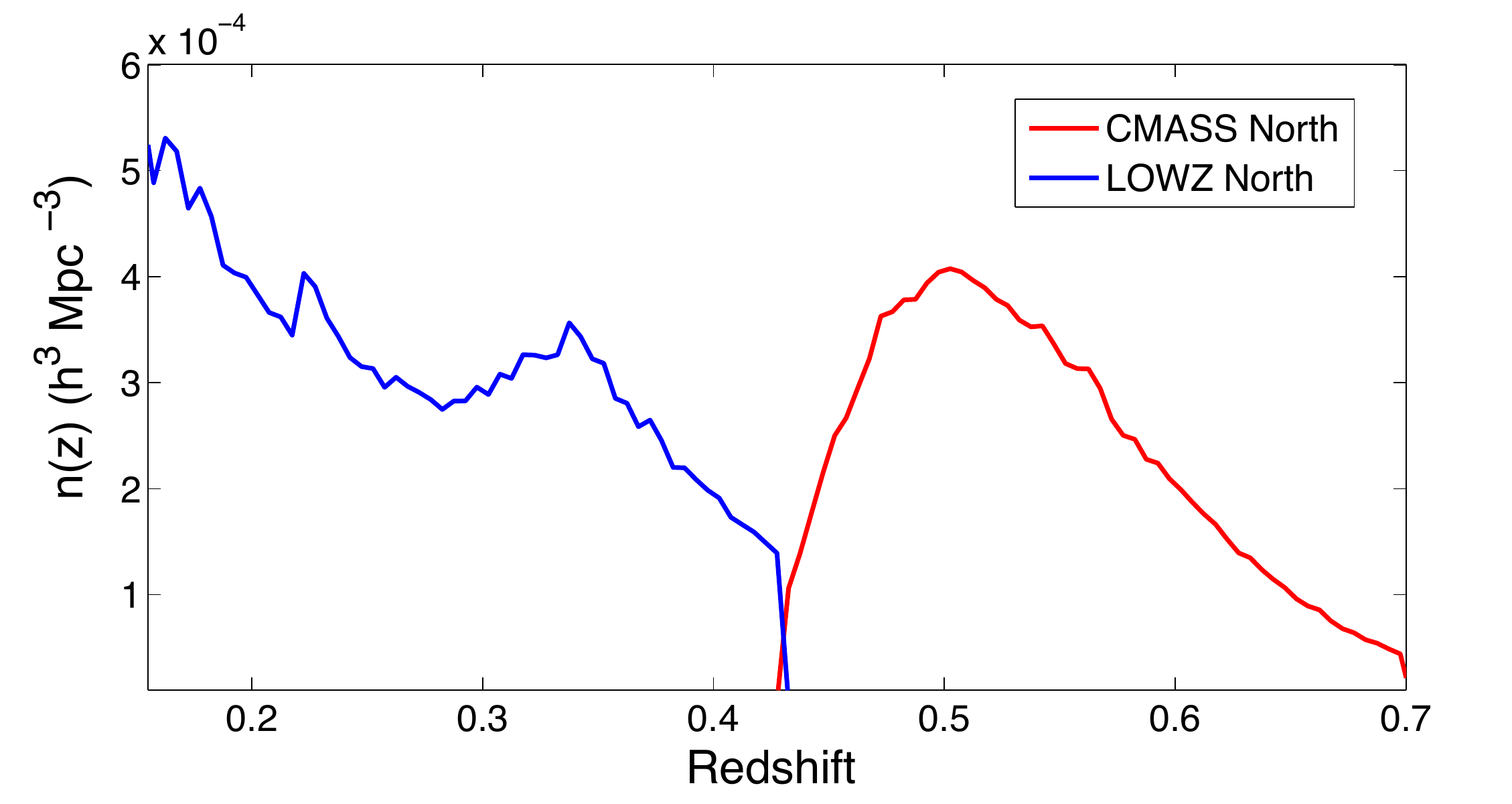}
\caption{The number density of galaxies as a function of redshift for both the North galactic cap of CMASS and LOWZ data.}
  \label{figure:nz}
\end{figure}
To empirically test the efficiency of reconstruction on distributions of galaxies with realistic masks, we make use of the PTHalo \citep{2013MNRAS.428.1036M} mocks created to match the Data Release 11 (DR11), Baryon Oscillation Spectroscopic Survey (BOSS) galaxy samples. BOSS \citep{2013AJ....145...10D} is part of the Sloan Digital Sky Survey III (SDSS; \citealt{2011AJ....142...72E}), a project that used the SDSS telescope \citep{2006AJ....131.2332G} to obtain imaging \citep{1998AJ....116.3040G} and spectroscopic \citep{2013AJ....146...32S} data, which was then reduced \citep{2012AJ....144..144B} to provide samples of galaxies from which clustering could be measured. Recent analyses of these data have benefited from having a large number of mocks, that have been used to estimate covariance matrices, and test methods. We use mocks created to match the angular mask corresponding to the galaxies included in Data Release 11 \citep{2014arXiv1401.4171M}.

BOSS measures redshifts for two galaxy samples, known as CMASS (which was selected to a approximately constant stellar mass threshold) covering $0.43\leq z\leq0.70$ and LOWZ (low redshift) sample with $0.15\leq z\leq0.43$ (further details about these samples, including the targeting algorithms used can be found in \citealt{2013arXiv1312.4877A}). A comparison of the redshift distribution of both samples is provided in Fig.~\ref{figure:nz}. The different redshift ranges mean that they cover different volumes, giving BAO measurements with different average precision. We will utilise samples of 600 mocks matched to the CMASS sampling, and 1000 LOWZ mocks. 

Because we use the PTHalos mocks extensively, we briefly review the process used to generate them. The method initially creates a matter field based on second order Lagrangian Perturbation Theory, displacing a set of tracer particles from their Lagrangian position by
\begin{equation}
\mathbf{\Psi} = \mathbf{\Psi^{\left(1\right)}} + \mathbf{\Psi^{\left(2\right)}},
\end{equation}
where the first order term is the Zel'dovich approximation and the second order term describes gravitational tidal effects 
\begin{equation}
\mathbf{\Psi\left(q\right)}^{\left(2\right)} \propto \sum_{i\neq j} \left({\frac{\partial\Psi_i^{\left(1\right)}}{\partial q_i} \frac{\partial\Psi_j^{\left(1\right)}}{\partial q_j}-\frac{\partial\Psi_j^{\left(1\right)}}{\partial q_i}\frac{\partial\Psi_i^{\left(1\right)}}{\partial q_j}}\right).
\end{equation}
Redshift space distortions are added to the mock galaxy distribution by modifying their redshifts according to the second order LPT peculiar velocity field in the radial direction.
The matter field is created in a single time-slice, rather than in a light cone, thus the growth rate and RSD signal are constant throughout the sample.  Halos are located with a Friends of Friends (FoF) algorithm, and halo masses calibrated to N-body simulations. The clustering of the halos is shown to be recovered to at least $\approx 10\%$ accuracy over the scales of interest for BAO measurements. The halos are populated with galaxies using a Halo Occupation Distribution (HOD) calibrated by the observed galaxy samples on small scales between $30\mpcoh$ and $80\mpcoh$. For the CMASS mocks, a non-evolving HOD was assumed, while the LOWZ mocks adopted a redshift dependant HOD \citep{2014arXiv1401.4171M}, with evolution introduced as a function of galaxy density. The mock galaxies are not assigned colour or luminosity.

The mocks are sampled to match the angular mask and redshift cuts of the survey data. Furthermore, to replicate some of the observational complications inherent in the BOSS survey, galaxies are sub-sampled to mimic missing galaxies caused by redshift failure, and close pairs - simultaneous spectroscopic observations are limited to objects separated by $>62^{\prime\prime}$. We weight mock galaxies using the FKP weighting scheme in \citet{1994ApJ...426...23F}, which we apply to calculate the displacement field, and to estimate the final clustering signal. The FKP weight is designed to optimally recover the over-density field given a sample with varying density, and is therefore appropriate to use for both measurements. We therefore apply a weight to each galaxy

\begin{equation}
  w = w_{FKP} \left(w_{cp} + w_{red} -1\right),
\end{equation}
where $w_{cp}$ and $w_{red}$ correct for the close-pairs and redshift failures respectively (see \citealt{2012arXiv1203.6594A} for further details), and $w_{FKP}$ is the FKP weight
\begin{equation}
w_{FKP} = \frac{1}{1 + \bar{n}\left(z\right)P_{0}},
\end{equation}
with fixed expected power spectrum $P_0 = 20,000h^{-3}\mathrm{Mpc}^3$, and average galaxy density $\bar{n}\left(z\right)$.

The clustering on intermediate scales is built up by interpolating between the small and large scales. Thus we see that galaxy displacements within the mocks will be formed from the structure growth (at second order) and a random component from the intra-halo velocities. Hence, they will provide a good test of reconstruction, although obviously, the intermediate scale clustering is not as accurate as it would have been had the mocks been calculated from N-body simulations, which we should bear in mind when interpreting our results. For simplicity in our analysis we use only the North Galactic Caps (NGC) of both sets; the CMASS NGC mocks each cover an effective area of 6,308 square degrees and the LOWZ NGC mocks have an effective area of 5,287 square degrees. Following previous work \citep{2013arXiv1312.4877A, 2014MNRAS.440.2222T}, we assume a linear bias value of $1.85$ for both samples which is calculated from the unreconstructed correlation function of the data. We use a linear growth rate of $f=0.74$ for CMASS and $f=0.64$ for the LOWZ sample. 

The DR11 PTHalo mocks have been used in a considerable number of previous BOSS analyses, measuring BAO (e.g. \citealt{2013arXiv1312.4877A, 2014MNRAS.440.2222T}), RSD (e.g. \citealt{2014MNRAS.439.3504S}), full fits to the clustering signal (e.g. \citealt{2014MNRAS.440.2692S}). We consider that they have therefore been extensively tested, and any limitations result from the method, as discussed above. 
\section{Measuring the BAO scale} \label{sec:fitting} 
In the following we will only consider measuring the BAO scale from spherically averaged 2-point clustering measurements. The monopole provides the majority of the important cosmological signal \citep{2013arXiv1312.4877A}, and thus is of most direct importance when testing the efficiency of reconstruction. Comparisons of BAO scale measurements made using either the monopole correlation function or monopole power spectrum have revealed a high degree of correlation \citep{2013arXiv1312.4877A, 2014MNRAS.440.2222T}. For simplicity, we therefore only consider fitting the power spectrum, as this requires significantly less computational effort to calculate.

The BAO scale is usually quantified with a dilation parameter $\alpha$ comparing the observed scale with that in the fiducial model used to measure the clustering statistic. For a measurement made from a monopole power spectrum to which all modes contribute equally, we define $\alpha$ as
\begin{equation}
  \frac{D_\textrm{V}}{r_d} = \alpha \left(\frac{D_\textrm{V}}{r_d}\right)_{fid},\label{eq:alpha}
\end{equation}
where $r_d$ is the comoving sound horizon at the drag epoch, and $D_V$ was defined in Eq.~(\ref{eq:Dv}). $\alpha$ can then be determined assuming that it linearly shifts the observed power spectrum monopole in wavelength. A value of $\alpha<1$ implies the acoustic peak appears at a larger scale than predicted by the fiducial cosmology. The goal of many modern galaxy surveys is to extract an unbiased value of $\alpha$ with a high level of precision. 

\subsection{Measuring the power spectrum}
To calculate the monopole power spectrum, we follow the standard procedure of \citet{1994ApJ...426...23F}, Fourier transforming the difference between a weighted galaxy catalogue and a weighted random catalogue with densities $\rho_{\rm gal}({\bf r})$ and $\rho_{\rm ran}({\bf r})$ respectively.
\begin{equation}
  F({\bf r}) = \frac{1}{N}\left[ \rho_{\rm gal}({\bf r}) w({\bf r})- \gamma\rho_{\rm ran}({\bf r})w({\bf r})\right],
\end{equation}
where $N$ is a normalisation constant for the integral
\begin{equation}
  N = \int{d^3r \rho_{\rm ran}^2({\bf r})w^2({\bf r})},
\end{equation}
$w\left({\bf r}\right)$ are the weights and $\gamma$ normalises the random catalogue, which is allowed to be denser than the galaxies
\begin{equation}
  \gamma=\frac{\sum \rho_{\rm gal}({\bf r}) w({\bf r})}{\sum \rho_{\rm ran}({\bf r}) w({\bf r})}.
\end{equation}
The spherically averaged measured power spectrum is defined as $P(k)=|F(k)|^2-F^2_{\rm shot}$, where 
\begin{equation}
  F^2_{\rm shot} = \left( 1 + \gamma\right) \frac{1}{N}\int{ d^3r \bar{n}\left(\mathbf{r}\right) w^2\left(\mathbf{r}\right)},
\end{equation}
is a shot-noise subtraction assuming the galaxies Poisson sample the underlying density field.

\begin{figure}
  \includegraphics[width=\linewidth]{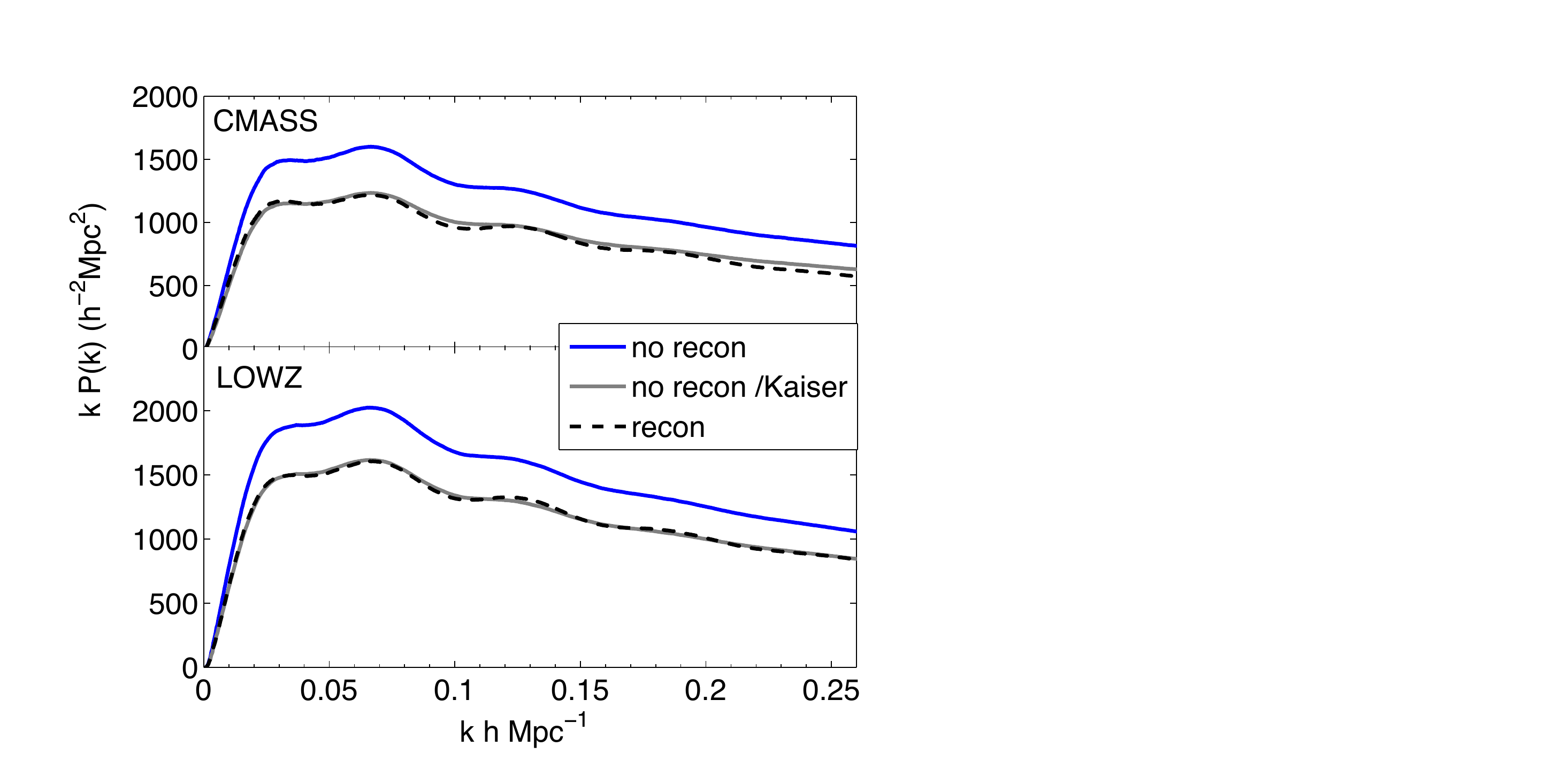} 
    \caption{Average power spectra of CMASS and LOWZ mocks pre and post reconstruction. The amplitude of the large scale power is decreased by the Kaiser factor ( $1+ 2/3\left(f /b \right) + 1/5 \left(f/b\right)^2$) when the linear RSDs are removed in the reconstruction process as shown by the dashed lines. The non-reconstructed power spectrum divided by the Kaiser factor is shown by the grey line.}
  \label{fig:power_pre_post}
\end{figure}
In our implementation of this routine, we calculate the power spectrum using the {\sc FFTW} package, on a $1024^3$ grid for a box of side length 3 Gpc. Example power spectra are presented in Fig.~\ref{fig:power_pre_post}, showing the pre-reconstruction power spectra compared to the post- reconstruction power spectra  (where the average power of the collection of mocks for each sample is shown). 
To show the amplitude has reduced by the expected amount for that redshift, we also include the pre-reconstruction power spectra divided by the linear Kaiser boost of $1+ 2/3\left(f /b \right) + 1/5 \left(f/b\right)^2$.

\subsection{Modelling the power spectrum}
To measure the baryon acoustic scale we follow \citet{2013arXiv1312.4877A} and fit our power spectrum measurement with a model consisting of a smooth broad-band term defined by a polynomial, multiplied by a model of the BAO signal which is rescaled by $\alpha$. The model power spectrum can be written 
\begin{equation}
P^{\rm m}\left(k\right) = P^{\rm smooth} \left(k\right)O^{\rm damp}\left(k/\alpha\right),\label{eq:powermodel}
\end{equation}
where the $P^{\rm smooth}\left(k\right)$ is the broadband power and $O\left(k\right)$ contains the BAO signal.
The linear power spectrum $P^{\rm lin}\left(k\right)$ is calculated using the {\sc Camb} package \citep{Lewis:2002ah}. Following \citet{2007ApJ...664..660E} and using the fitting formula of \citet{1999ApJ...511....5E} a model of the \mydoubleq{De-wiggled} smooth power spectrum $P^{\rm sm,lin}\left(k\right)$ is used to decouple the linear BAO feature $O^{lin}$ from the linear power spectrum,
\begin{equation}
P^{\rm lin}\left(k\right) = P^{\rm sm,lin}\left(k\right)O^{\rm lin}\left(k/ \alpha\right).
\end{equation}
To account for non-linear structure formation, the linear BAO signal is damped 
\begin{equation}
O^{\rm damp}\left(k/\alpha \right) = \left(O^{\rm lin}\left(k/\alpha \right)-1 \right) e^{-k^2\Sigma^2_{\rm nl}/2} +1.
\end{equation}
The damping scale $\Sigma_{\rm nl}$ is fixed using values derived from the average damping recovered from the mocks pre/post reconstruction. We use; CMASS, pre-reconstruction $8.3\mpcoh$, post-reconstruction $4.6\mpcoh$; and LOWZ, pre-reconstruction $8.8\mpcoh$, post-reconstruction $4.8\mpcoh$.

The smooth broadband part of the power spectrum is calculated using a model constructed with 5 polynomial terms $A_i$ and a multiplicative term $B_p$ that accounts for large-scale bias \citep{2013arXiv1312.4877A,2013MNRAS.428.1116R}
\begin{equation}
P^{\rm sm}\left(k\right) = B_p^2 P\left(k\right)^{\rm sm,lin} + A_1k + A_2 + \frac{A_3}{k} + \frac{A_4}{k^2} + \frac{A_5}{k^3}.
\end{equation}

To replicate the effects of the survey geometry, a window function ($|W(k)|^2$) is constructed from the normalised power spectrum of the random catalogue as shown in \citet{2007MNRAS.381.1053P}. This is convolved with the model power spectrum over $0 < k_i< 2\hompc$.

\begin{figure}
  \includegraphics[width=\linewidth]{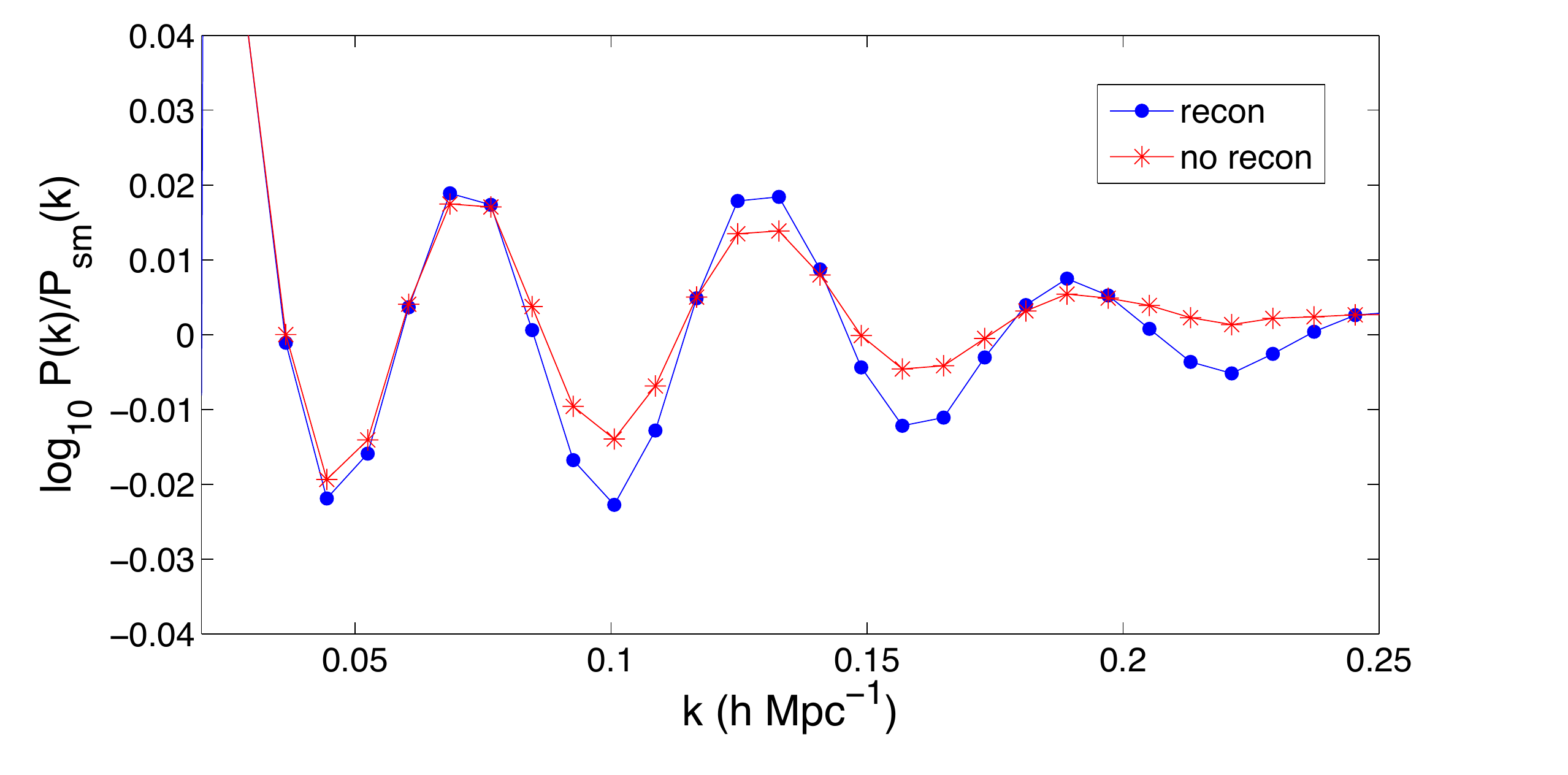} 
    \caption{Average of 600 CMASS mock power spectra divided by the no-wiggle model, pre reconstruction is shown by the red line and post reconstruction is the blue line. The plot shows how the oscillations are less damped post reconstruction. The discreteness is a result of the power spectrum binning choice.}
  \label{fig:wiggles}
\end{figure}

A plot showing the average pre and post reconstruction power spectra of the 600 CMASS catalogues divided by the smooth model is shown in Fig.~\ref{fig:wiggles}. It is clear that the reconstruction process has reduced the damping of the BAO on small scales.

\subsection{Fitting the BAO scale}

For each mock analysed we calculate a likelihood surface for $\alpha$, covering the range from $0.8<\alpha<1.2$ with separation of $\Delta\alpha = 0.002$. At each point we marginalise over the polynomial parameters, and calculate the likelihood assuming that all parameters were drawn from a multi-variate Gaussian distribution. 

We characterise how well the reconstruction algorithm works by comparing the pre and post reconstruction $1\sigma$ errors, calculated by marginalising over the likelihood surface, which we call $\sigma_{\alpha, \mathrm{pre}}$ and $\sigma_{\alpha, \mathrm{post}}$ respectively. From each set of mocks, we also calculate the mean values of these errors $\langle \sigma_{\alpha,{\rm pre/post}}\rangle$, and the standard deviation of the distribution of marginalised best-fit $\alpha$ values,  $S_{\alpha,{\rm pre/post}}$ for comparison. 

To account for a different number of LOWZ and CMASS mocks we include a correction on the errors to compare samples (as described in \citealt{2014MNRAS.439.2531P}). There are two corrections, the first follows from our method of estimating the inverse covariance matrix leading to a bias that can be corrected by a renormalisation of the $\chi^2$ value. The second comes from the propagation of errors within the covariance matrix which can be corrected with different multiplicative factors applied directly to the variance of the sample, and to the recovered $\sigma_{\alpha}$.
\begin{figure*}
    \centering
    \resizebox{0.8\columnwidth}{!}{\includegraphics{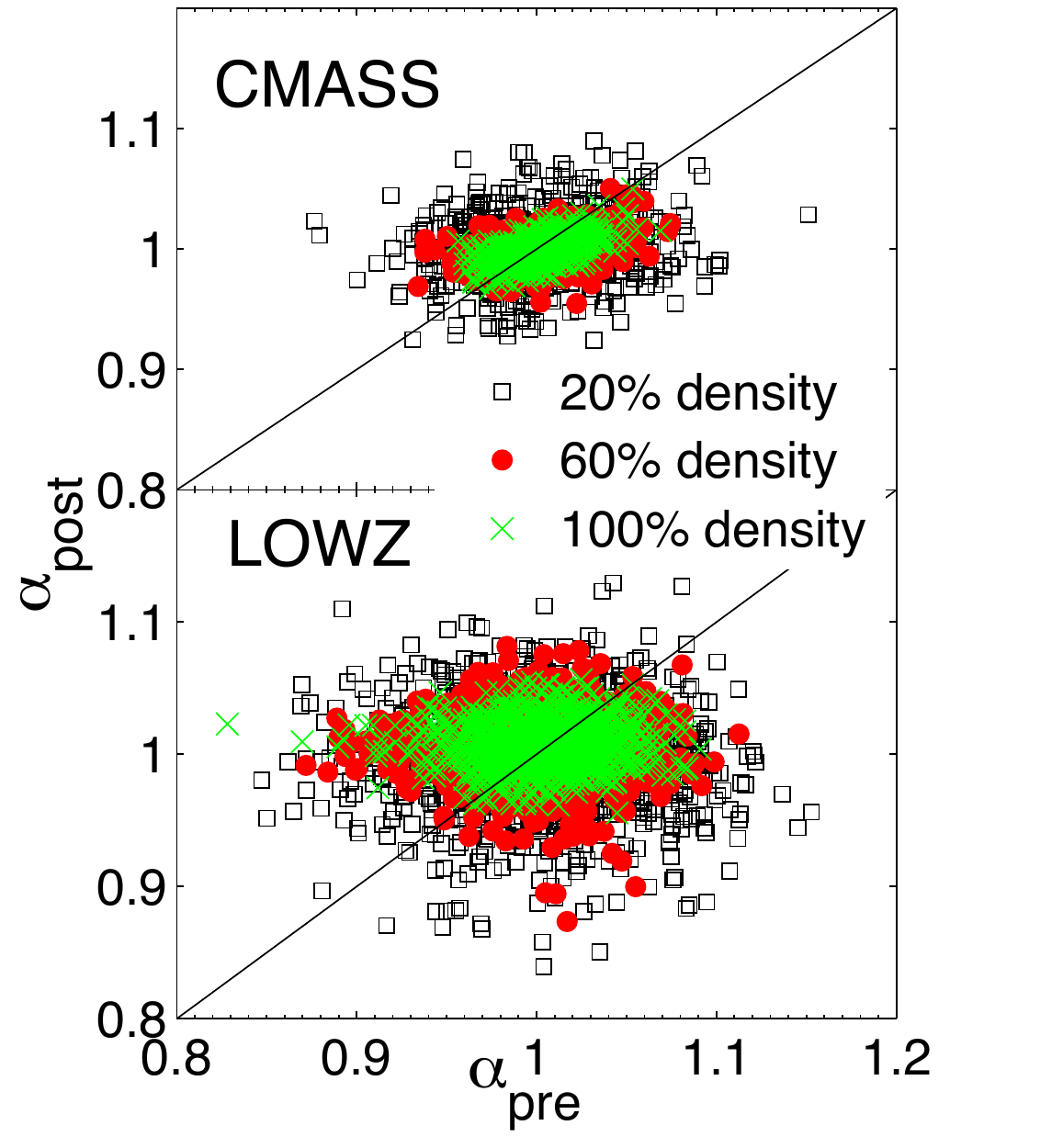}}
    \hspace{1cm}
    \resizebox{0.8\columnwidth}{!}{\includegraphics{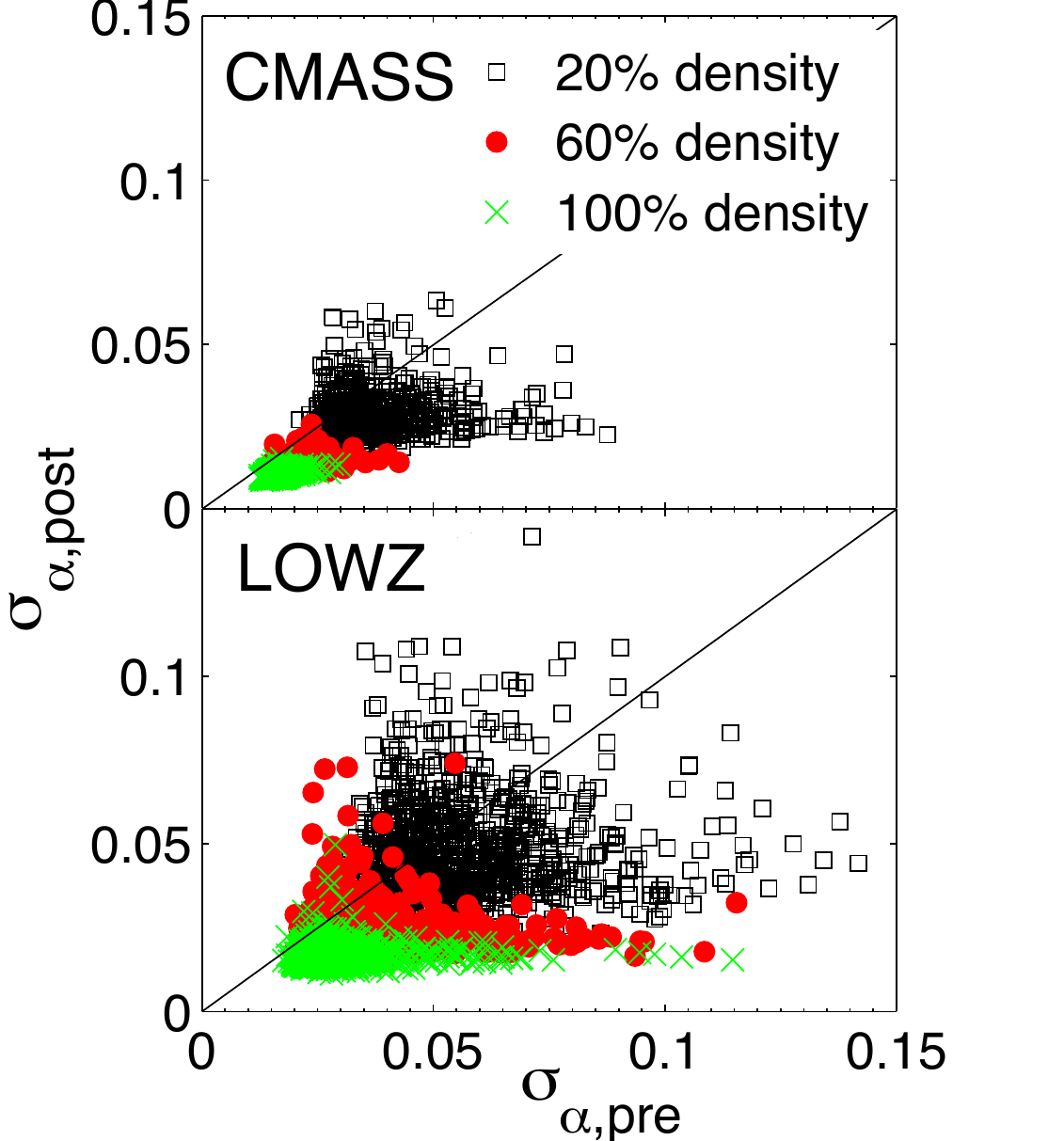}}
    \caption{Recovered $\alpha$ (left) and $\sigma_{\alpha}$ (right) values from power spectrum fits of both CMASS and LOWZ samples. The pre reconstruction values are on the x-axis and the post reconstruction values on the y-axis. The black squares indicate samples cut to 20\% of their original density, the red points indicate 60\% of the original density and the green crosses are the samples at 100\% density. Clearly the scatter in both sets of plots is reduced for both pre and post reconstruction measurements as the density of the sample is increased. The CMASS samples show less scatter than the LOWZ samples in both graphs and the recovered errors are smaller. Reconstruction clearly reduces the recovered $\sigma_{\alpha}$ values on average in all of the samples although the fraction of mocks that show improvement increases with sample density.}
  \label{fig:density1}
\end{figure*}

\section{Change in effectiveness with survey density}  \label{sec:density}
  \begin{table*}
 \centering
 \begin{minipage}{165mm}
  \caption{BAO scale errors recovered for different survey densities from the LOWZ and CMASS mocks.}
    \label{table:BAO_density} 
  \begin{tabular}{@{}lllllllllc@{}}
  \hline
Sample & Density(\%)   &  V$_{\mathrm{eff}} (h^{-3}$Gpc$^{3}$) & 
$\langle\alpha_{\mathrm{post}}\rangle$ &   $\langle \sigma_{\alpha, \mathrm{post}}\rangle$ & $S_{\alpha, \mathrm{post}}$ & $\langle\alpha_{\mathrm{pre}}\rangle$ &  $\langle \sigma_{\alpha, \mathrm{pre}}\rangle$ &  $S_{\alpha, \mathrm{pre}} $ & \%  with $\frac{\sigma_{\alpha,\mathrm{post}} }{\sigma_{\alpha, \mathrm{pre}}}<1$\\
 \hline
 CMASS & 100  & 1.12&  0.9998  & 0.0112 & 0.0109     & 1.0032    & 0.0173   &     0.0172 & 100 \\
                & 80     &  0.97&  1.0005 & 0.0130 &  0.0125     &1.0038     &0.0185    &     0.0185 & 100\\
                & 60     &  0.78&  0.9997 & 0.0141 & 0.0140    &1.0036     &0.0212   &    0.0214   & 99.5 \\ 
                & 40     & 0.54&  0.9994   & 0.0182 &  0.0182     &1.0035     &0.0237   &    0.0244 & 94.3\\  
                & 20     &  0.23&  1.0009  & 0.0303 & 0.0287     &1.0037     &0.0384    &    0.0363 & 79.0 \\  
\hline
LOWZ & 100 &0.52& 0.9997 & 0.0169 & 0.0157  &  1.0035  & 0.0302  & 0.0308 & 99.2  \\
           & 80   &0.47& 0.9992  &  0.0208 & 0.0216  &  1.0031  & 0.0323  & 0.0334 &  93.3\\
           & 60   &0.39& 1.0041 & 0.0236  &0.0254 &  1.0006  & 0.0348  & 0.0344    & 90.3\\
           & 40   &0.29& 1.0014  &  0.0304 & 0.0314  &  1.0014  & 0.0418  & 0.0406  & 83.0 \\
           & 20   &0.14& 0.9959  &  0.0493 & 0.0425  &   1.0008  & 0.0579 & 0.0499  & 65.8\\
\hline
\end{tabular}
\end{minipage}
\end{table*}
\begin{figure*}
    \centering
    \resizebox{0.46\textwidth}{!}{\includegraphics{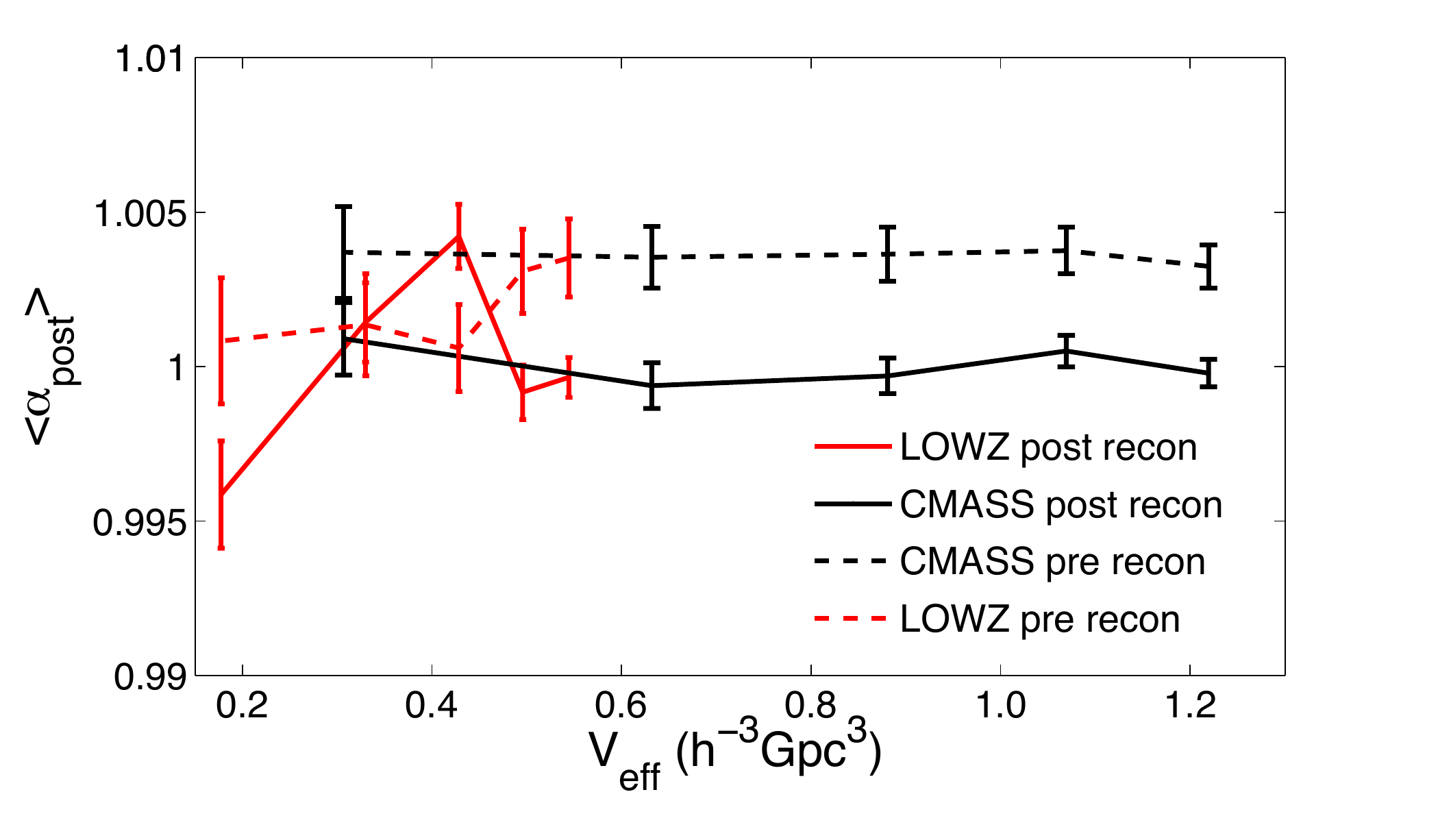}}
    \hspace{1cm}
    \resizebox{0.46\textwidth}{!}{\includegraphics{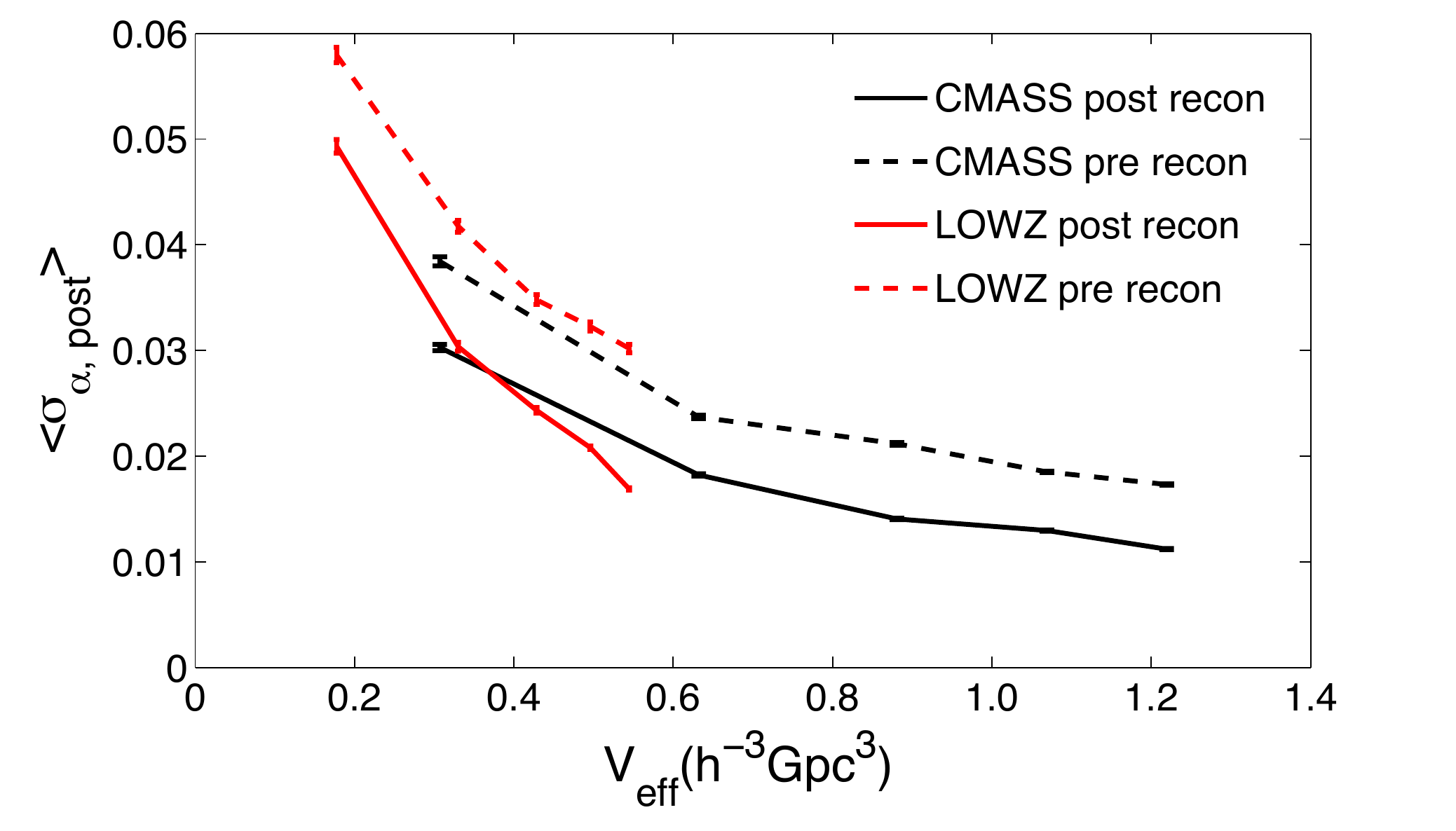}}
    \caption{ Recovered $\langle\alpha\rangle$ (left)  and $\langle\sigma_{\alpha}\rangle$ (right) from power spectrum fits for CMASS and LOWZ as a function of effective volume. The  $\langle\alpha\rangle$ values are consistent in the range of CMASS subsamples. The value is biased high pre reconstruction (black dashed line), and the bias is removed by reconstruction such that the values are consistent with 1 (black full line).  The pre reconstruction LOWZ sample (red dashed line) shows no bias in $\langle\alpha\rangle$ pre reconstruction for sub samples at a lower effective volume. When the effective volume is increased the bias in the pre reconstruction $\langle\alpha\rangle$ measurement becomes apparent and is removed post reconstruction (red full line). The average $\sigma_{\alpha,post}$ values are clearly reduced with increasing effective volume both pre and post reconstruction for both samples. }
  \label{fig:density_Veff}
\end{figure*}
Although reconstruction is a non-local process, there are only mild correlations between regions separated on large scales of order of the survey size, such that we expect the galaxy number density to drive the effectiveness of reconstruction rather than the survey volume. Increasing the galaxy density reduces the shot noise in measurements of the displacement field and as a result we would expect the reconstruction to be more efficient. In this section we quantify this effect by comparing the pre and post reconstruction errors after subsampling the galaxy catalogues to match $20\%,40\%, 60\%$, $80\%$ and $100\%$ of the original density keeping the same relative redshift distribution. As a result of tests carried out in Section~\ref{sec:method} we use a smoothing length of 15$\mpcoh$ for the CMASS sample and 10$\mpcoh$ for the LOWZ sample.

In addition to reconstruction, the error on post-reconstruction BAO-scale measurements depends on the volume through an interplay with the survey density, in such a way that the error decreases as the survey density and volume increase. The combination can be characterised by an effective volume  \citep{1994ApJ...426...23F,1997PhRvL..79.3806T},
\begin{equation}  
\label{eq:Veff}
  V_{\mathrm{eff}} \left(k\right) \equiv \int{\left[ \frac{ \bar{n} \left( \mathbf{r} \right) P_{s,0}}{1 + \bar{n} \left( \mathbf{r} \right) P_{s,0}} \right]^{2}}d^3r,
\end{equation}
which also depends on the power spectrum amplitude in redshift space, which we denote $P_{s,0}$. In the following we use the measured value at $k\approx 0.1\hompc$. The power spectrum error is inversely proportional to the square root of the effective volume for a given sample. We expect the BAO precision without reconstruction to depend on this and the degree of BAO damping. We choose to plot our measurements of BAO scale errors against effective volume, even though we only change the galaxy density for each sample. This allows us to simultaneously present LOWZ and CMASS results against a consistent baseline. We compare the improvement in error due to reconstruction for each sample which being a relative measurement can be directly compared to the average survey density. We also compare the relative improvement of reconstruction against $\bar{n}P_{s,0}$. This allows us to separate the efficiency of reconstruction from the amplitude of the clustering signal.

Fig.~\ref{fig:density1} compares pre vs post-reconstruction $\alpha$ and $\sigma_{\alpha}$ on a mock by mock basis. These plots show points for a subset of the revised density catalogues, clearly showing that increasing the density of the survey reduces the scatter in $\alpha$ and $\sigma_{\alpha}$. The distribution of $\alpha$ values in both samples follows a locus with shallower gradient than the solid line showing that, on average, post-reconstruction values are closer. We see a corresponding improvement in the values of $\sigma_{\alpha}$ where all points that fall below the solid line indicate a reduction in error post reconstruction. The $\sigma$ values extracted from the CMASS measurements are clearly smaller than the LOWZ values both pre and post reconstruction. As the density of a sample is increased both $\sigma$ values and their scatter decreases.

The $\langle \alpha \rangle$ and $\langle \sigma_{\alpha}\rangle$ values recovered from each set of mocks pre and post-reconstruction are collated in Table~\ref{table:BAO_density}. 
Predictions in \citet{2007ApJ...664..675E} suggest that non-linear structure formation induces a small bias in the acoustic scale measured in the galaxy distribution of the order of 0.5\% . 
Pre-reconstruction, the CMASS sample shows a small bias in the mean recovered $\alpha$ away from the true value $\alpha=1$. The bias is consistent, between 0.3\% and 0.4\% high for the range of densities analysed, according to predictions. 
Tests on high resolution simulations suggest that this bias should be reduced by reconstruction to 0.07\% - 0.15\% \citep{2011ApJ...734...94M}. The correction due to reconstruction is shown to be a consequence of reducing the amplitude of mode coupling terms in the density field apparent at low redshift \citep{PhysRevD.79.063523}. 
Post reconstruction the bias reduced in all of the CMASS samples. At 100\% density the bias is reduced to 0.02\% high of the true value, below the statistical uncertainty on $\langle\alpha\rangle$ of 0.05\%. At all densities the post-reconstruction CMASS $\langle\alpha\rangle$ are within $1\sigma$ of the true value and are significantly lower than the error on any one realisation. The standard deviation of $\alpha$ values for a set of mocks are 
consistent with the $\langle \sigma_{\alpha} \rangle$ values confirming the validity of our likelihood calculations.
Pre-reconstruction, the lower density (20\%, 40\% and 60\%) LOWZ $\langle\alpha\rangle$ values are within 0.1\% of 1. There is weak evidence that the LOWZ bias increases with $V_{\rm eff}$, and at 100\% density, the bias in the LOWZ sample is increased to ~0.4\% inline with the pre reconstruction CMASS samples. This suggests that the low bias in the low density samples is a \mysingleq{lucky} coincidence, a consequence of under-sampling the density and losing small scale information. At a higher redshift, the galaxies in the CMASS mocks are not as tightly clustered which may explain why this effect is only seen in the LOWZ sample. 
Post reconstruction, the bias in the measurement of $\langle\alpha\rangle$ increases from 0.08\% to 0.4\% high in the 20\% sample, remains the same for the 40\% sample and increases from 0.06\% to 0.4\% high for the 60\% sample. Thus for these low density LOWZ samples, reconstruction fails to move the average $\langle\alpha\rangle$ values closer to 1. If the initial recovered $\langle\alpha\rangle$ values are not as expected (ie biased away from 1) due to high shot noise in the galaxy density field, it is unlikely that using this distribution of galaxies to measure the Lagrangian displacement field will enable reconstruction to accurately correct the density field.
However, as the density of the LOWZ sample is increased, the bias values fall in line with predictions. In these cases reconstruction reduces the bias in the recovered $\langle\alpha\rangle$ values. At 100\% density, the pre-reconstruction value is biased by 0.4\% high, this is reduced to 0.03\% low post reconstruction within the statistical uncertainty on one measurement of 0.05\%, at 80\% density the bias is reduced from 0.3\% high to 0.08\% low, these results are consistent with the CMASS results and predictions.

Graphs of $\langle\alpha\rangle$ and $\langle\sigma\rangle$ for all density subsets of CMASS and LOWZ are shown in Fig.~\ref{fig:density_Veff}. The CMASS $\langle\alpha\rangle$ values are very consistent pre and post reconstruction. The LOWZ results only become consistent with the CMASS results and predictions above an effective volume of 0.5 $\gpcohV$. The $\langle\sigma\rangle$ show a clear reduction with effective volume for both samples both pre and post reconstruction. The LOWZ errors are higher than the CMASS pre reconstruction due to the more advanced non-linearities in the density field. However, as the effective volume is increased, the LOWZ post reconstruction error rapidly decreases and surpasses the CMASS error suggesting that for a given effective volume, reconstruction works harder for the lower redshift sample.

\begin{figure}
  \includegraphics[width=\linewidth]{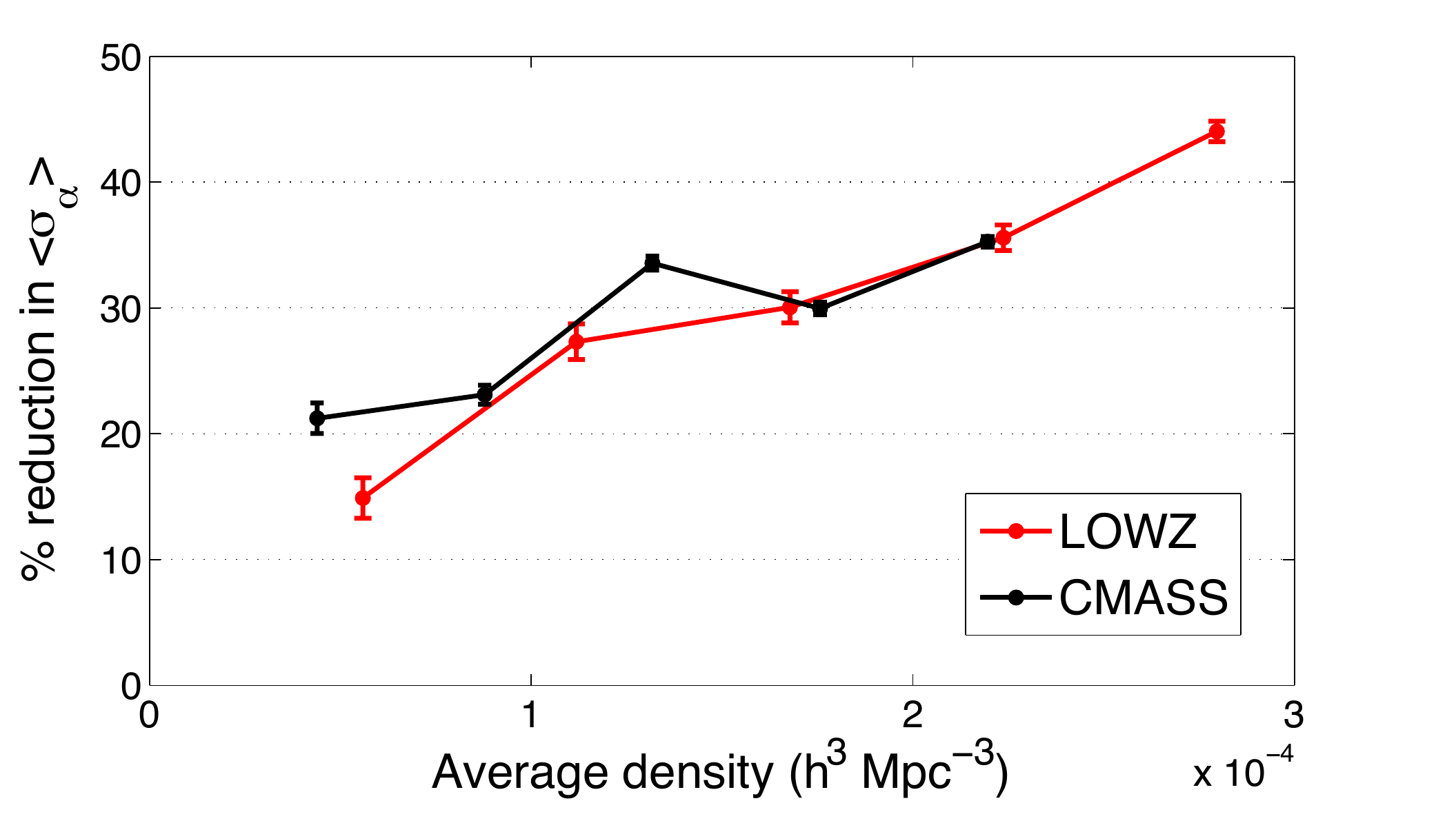} 
    \caption{Percentage improvement, $100\times(1 - \langle\sigma_{\alpha,post}\rangle /\langle \sigma_{\alpha, \mathrm{pre}}\rangle) $, on $\sigma_{\alpha}$ recovered after reconstruction for both CMASS (black line) and LOWZ (red line) samples as a function of $\bar{n}$. The improvement clearly increases with the average survey density in both cases. }
  \label{fig:relIMP}
\end{figure}

We quantify how effective our reconstruction algorithm is by comparing the percentage reduction in $\langle\sigma_{\alpha}\rangle$ before and after applying the algorithm. Fig.~\ref{fig:relIMP} shows the improvement $100\times (1 -\langle \sigma_{\alpha,{\rm post}}\rangle / \langle\sigma_{\alpha, \mathrm{pre}}\rangle) $ as a function of $\bar{n}$. Both sets of results show that the efficiency of reconstruction is increased as the density of the survey is increased. The 3rd point on the CMASS curve representing the 60\% density sample in Fig~\ref{fig:relIMP} is an outlier and does better than the 80\% density sample, although its absolute error is larger. 
For the LOWZ sample the efficiency drops more rapidly once the galaxy density is below ~1$\times 10^{-4}\hompcV$. 
However, the CMASS sample seems to show a constant decline in the efficiency of reconstruction with the reduction in survey density. There is no suggestion that the efficiency will asymptote at an optimal density. Performing a simple linear fit on the data we find that the fractional reduction in error, $1 -\langle \sigma_{\alpha,{\rm post}}\rangle / \langle\sigma_{\alpha, \mathrm{pre}}\rangle \approx 1000\bar{n} + 0.13$. This suggests that for a reduction in error of $50\%$, the survey density should be approximately 4$\times 10^{-4}\hompcV$.

\begin{figure}
  \includegraphics[width=\linewidth]{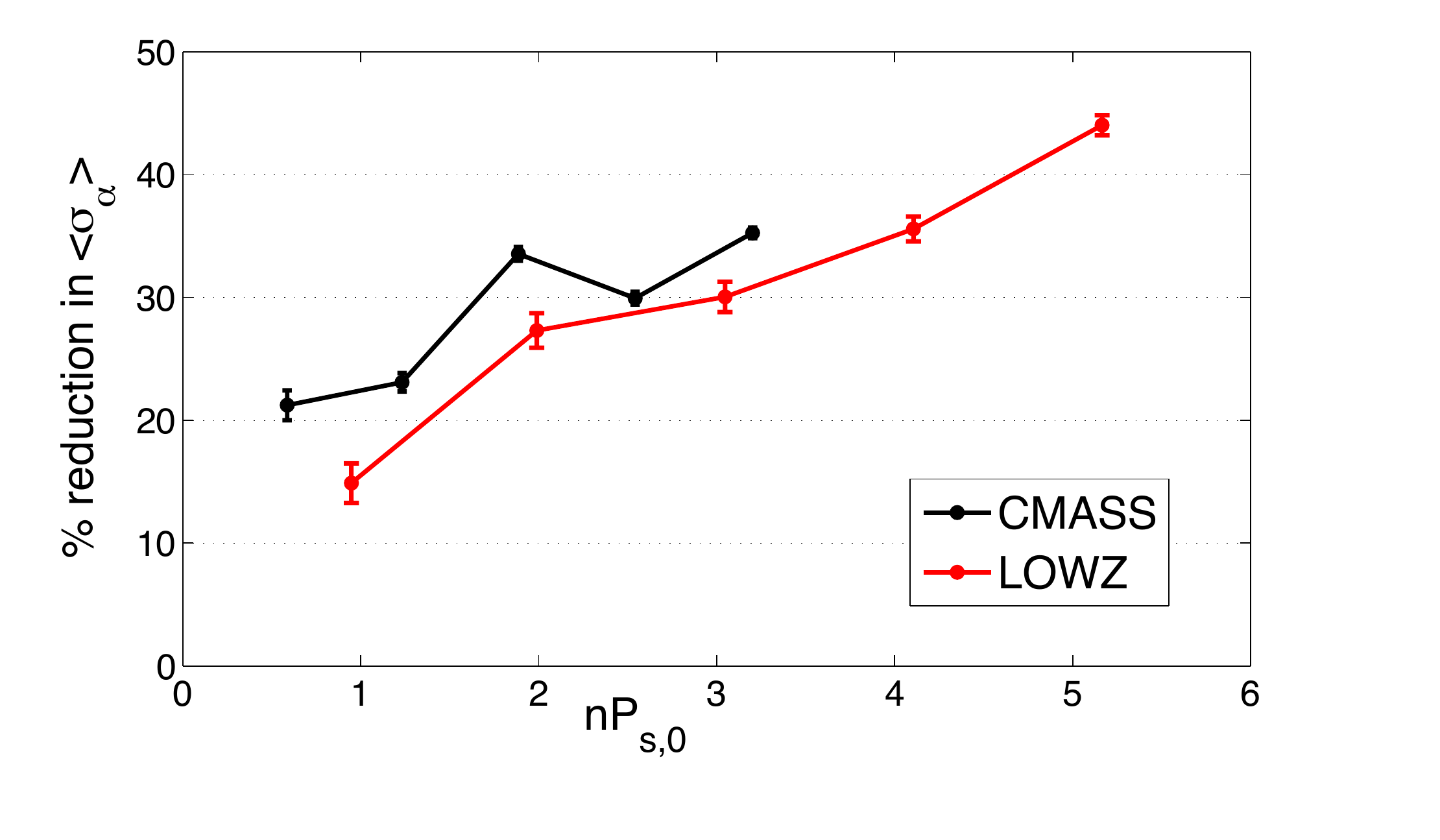} 
    \caption{Percentage improvement, $100\times(1 - \langle\sigma_{\alpha,post}\rangle /\langle \sigma_{\alpha, \mathrm{pre}}\rangle) $, on $\sigma_{\alpha}$ recovered after reconstruction for both CMASS (black line) and LOWZ (red line) samples as a function of $\bar{n}P{s,_0}$. There is a clear trend of improvement as $\bar{n}P_{s,0}$ is increased in both cases, although LOWZ does slightly worse than CMASS for a given $\bar{n}P_{s,0}$.}
  \label{fig:improve_vs_np}
\end{figure}

In Fig.~\ref{fig:improve_vs_np}, the effectiveness of reconstruction is compared to the the $\bar{n}P_{s,0}$ quantity, thus removing the clustering strength dependence from the comparison. The two curves show a clear trend of increasing efficiency with $\bar{n}P_{s,0}$. The higher $P_{s,0}$ value of the LOWZ sample moves the curve to the right compared to the CMASS curve. Thus compared to Fig.~\ref{fig:relIMP} the LOWZ sample does not do as well as the CMASS sample for a given $\bar{n}P_{s,0}$. This suggests that for a given sample, a higher clustering signal amplitude would increase the effectiveness of reconstruction. This is expected as at low redshifts where the clustering is more evolved, there is a greater non-linear contribution to the density field to remove, and the density perturbations that source the Lagrangian displacement fields are larger.

Histograms of the $\alpha$ and $\sigma_\alpha$ values recovered from the mocks for CMASS, LOWZ pre and post-reconstruction are shown in Fig.~\ref{fig:CMASSdensityHIST} and~\ref{fig:LOWZdensityHIST} in Appendix~\ref{app:hist}.

%
\section{Change in effectiveness near edges}
\label{sec:edges}
\begin{figure*}
    \centering
    \resizebox{0.46\textwidth}{!}{\includegraphics{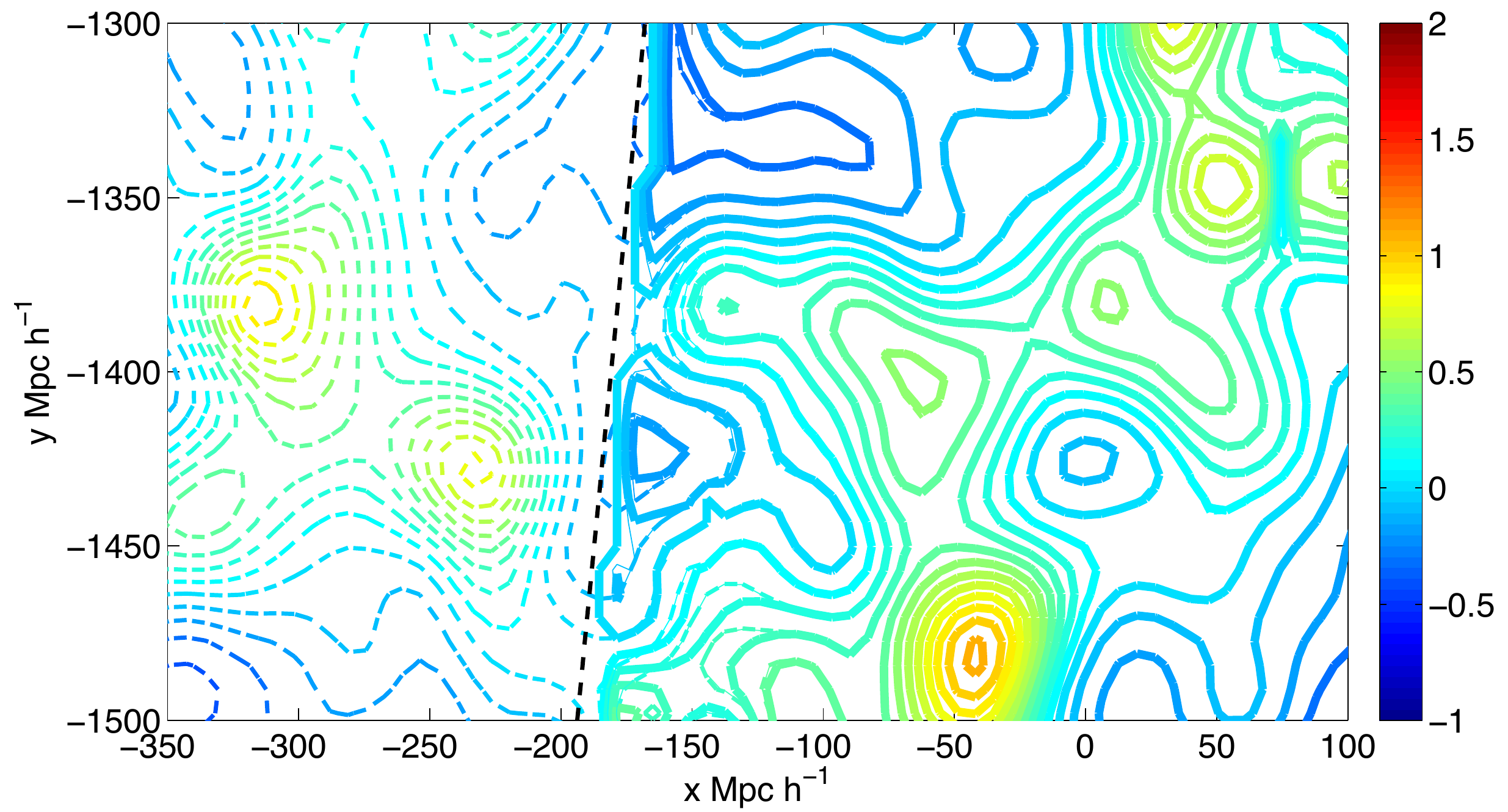}}
    \hspace{1cm}
    \resizebox{0.46\textwidth}{!}{\includegraphics{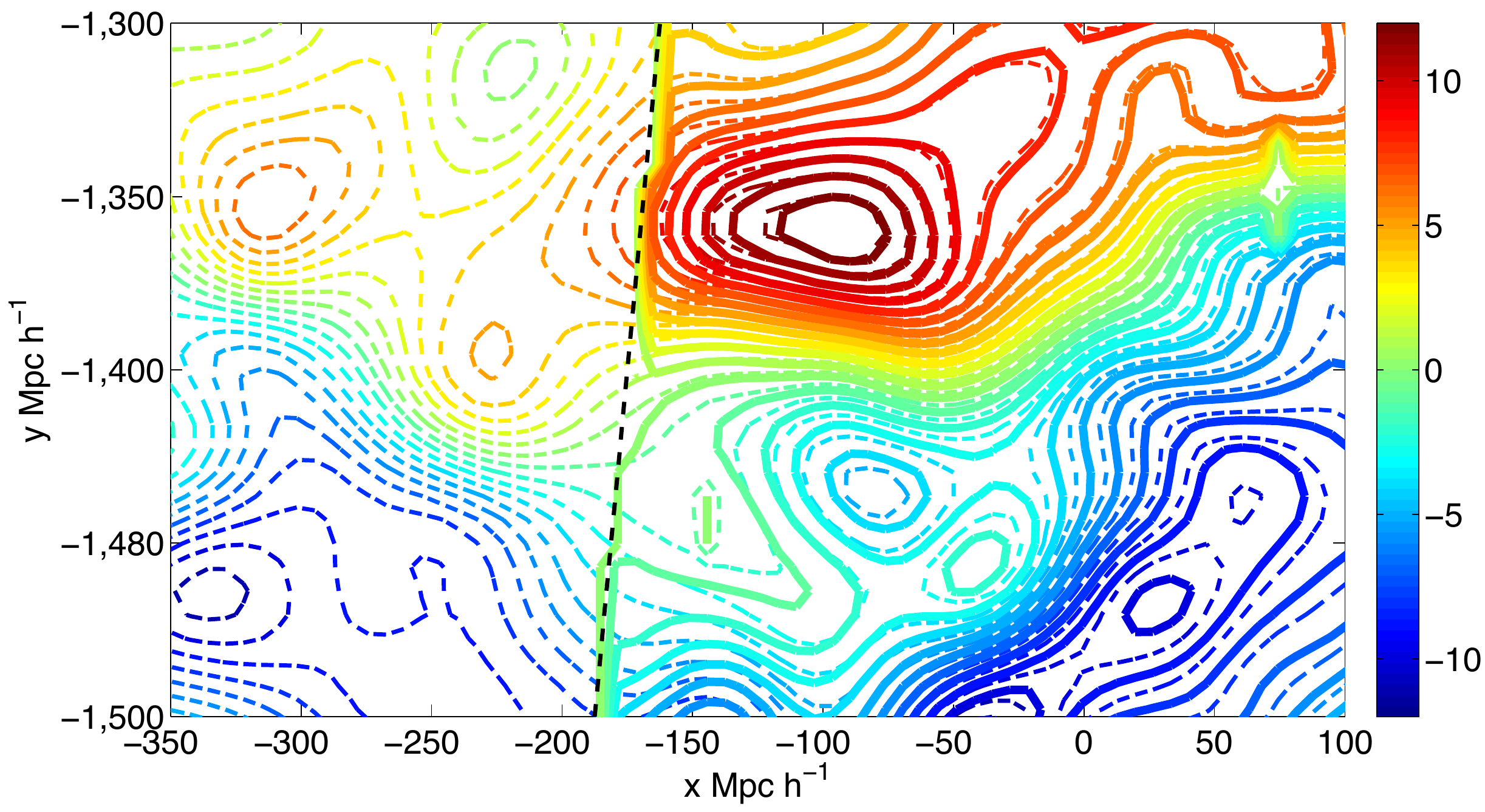}}
    \caption{ The left figure shows the smoothed overdensity field, the right hand figure shows the amplitude of the Lagrangian displacement field in the x direction. The dashed lines show the original fields and the full lines show the field recovered using only the information to the right of the dashed black line.}
  \label{fig:fields}
\end{figure*}
\begin{figure}
\minipage{0.44\textwidth}
    \resizebox{\textwidth}{!}{\includegraphics{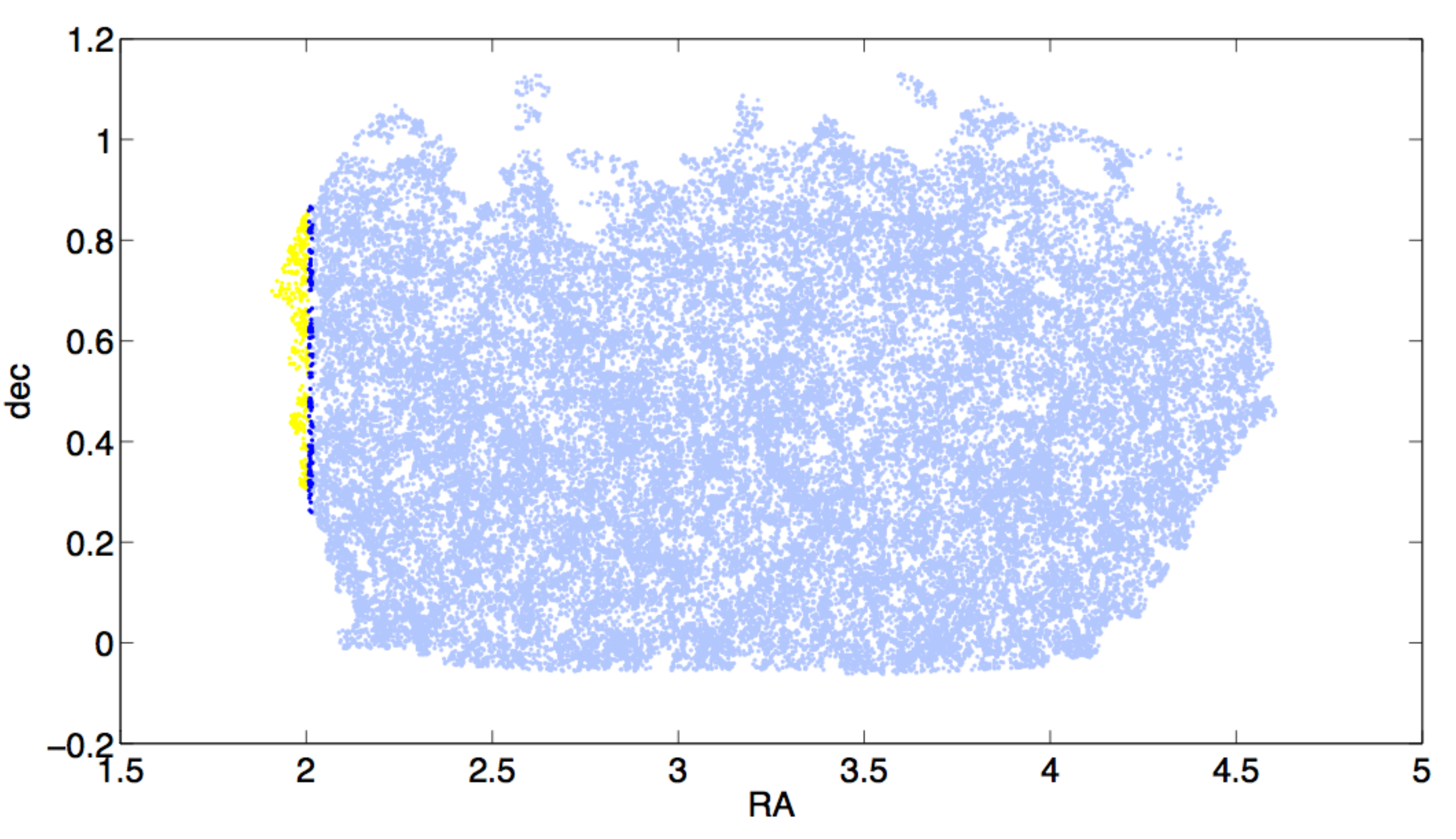}}
\endminipage\hfill
\minipage{0.44\textwidth}
    \resizebox{\textwidth}{!}{\includegraphics{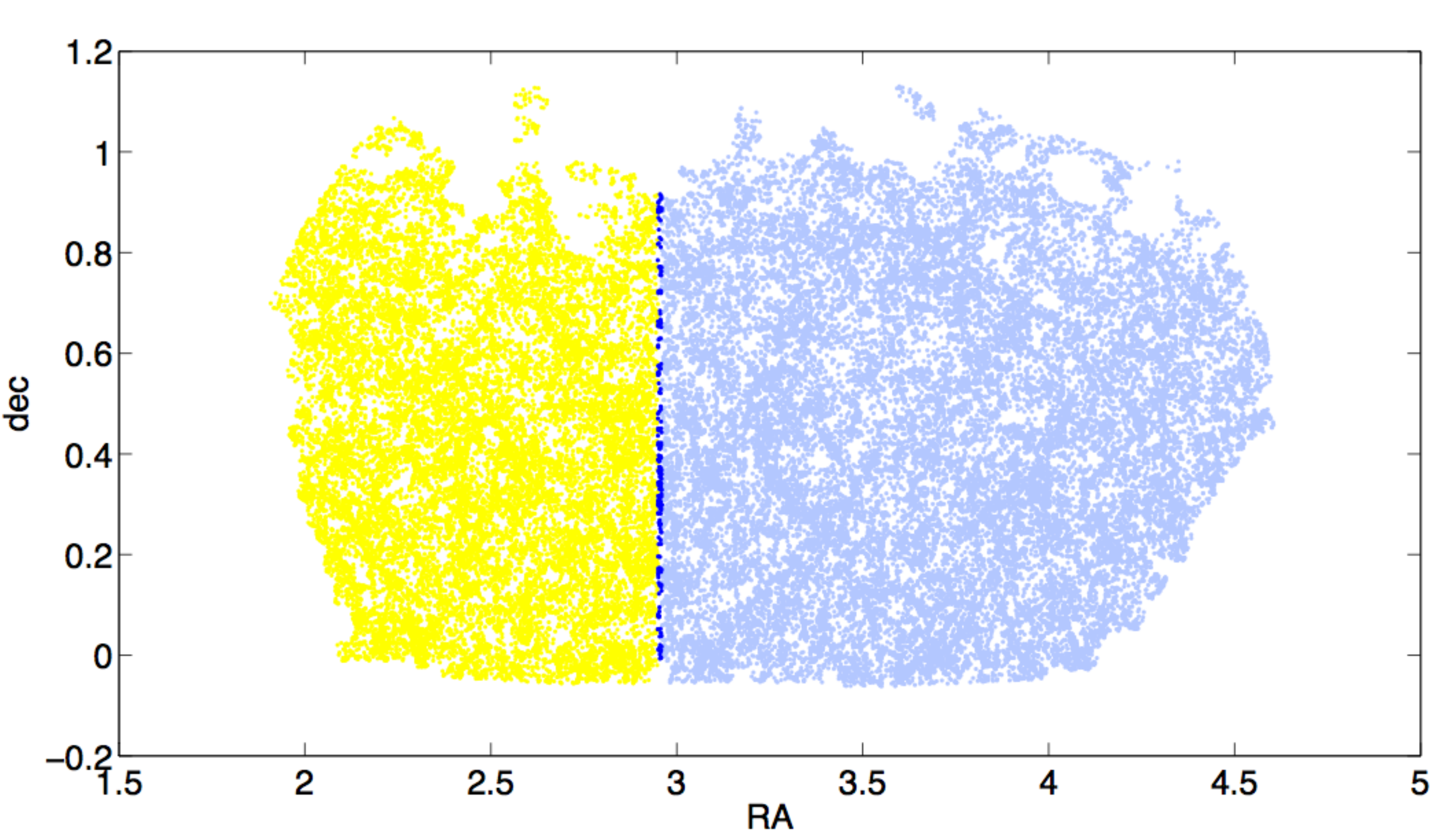}}
\endminipage\hfill
\minipage{0.44\textwidth}%
  \resizebox{\textwidth}{!}{\includegraphics{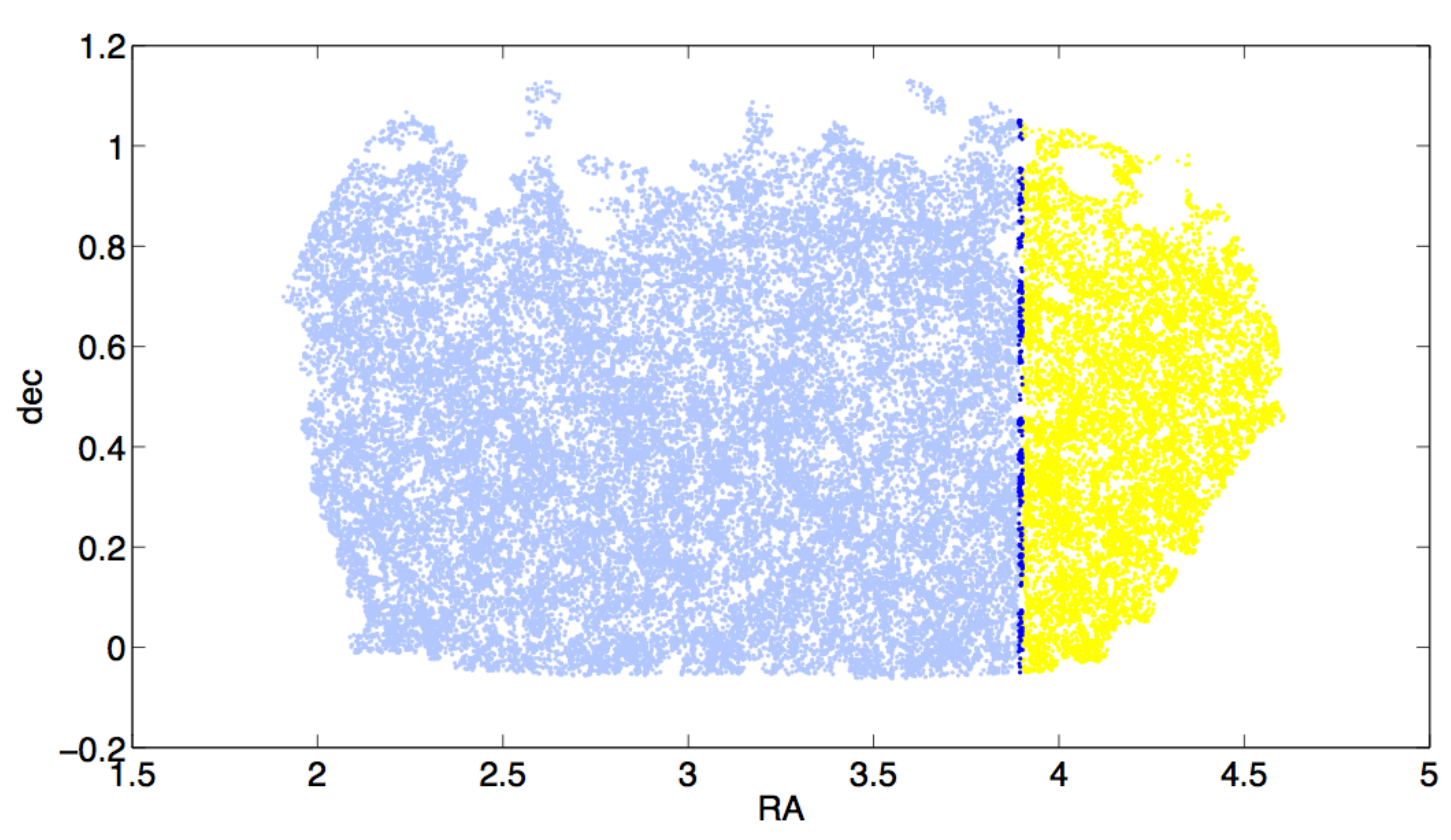}}
\endminipage
\caption{Plot showing how we impose edges on all galaxies within the sample. In each panel, only the dark blue galaxies are reconstructed, and this reconstruction only uses information from the galaxies shown in light and dark blue. The figures depict stripes 10, 100 and 190 respectively, out of the 257 stripes that we split the sample into. Once we have applied reconstruction for each of the 257 stripes, and measured the galaxy and random displacements in that stripe, we stitch the galaxy and random catalogues back together to give a full sample, reconstructed as if all galaxies lie close to an angular boundary.}
  \label{fig:stripes}  
\end{figure}

At a survey boundary, due to a reduction in information describing the surrounding overdensity, we expect reconstruction to be less efficient. Although we do not expect this ``edge effect'' to be substantial for the CMASS sample, which has a large volume to edge ratio, we attempt to quantify it in this section, as it will be of interest for future surveys. The effect of an artificial edge is shown in Fig.~\ref{fig:fields}, which shows a thin redshift slice through one mock where a survey boundary (the black line) has been artificially imposed. Dashed contours show the full density (left) and displacement field amplitude in one dimension (right) calculated using the full sample, and the solid contours show the result of cutting along the boundary. After excising the information to the left of the dashed black line, we see that both density and displacement fields are damped at the boundary, and the displacement field is mildly distorted on larger scales. This matches expectation: The displacement estimated for a galaxy positioned near an edge of the survey will not be influenced by anything beyond the boundary.

Although the reconstruction process is non-local it is expected that the influence of the edges on larger scales to be small (such as seen in the distortions in the Lagrangian displacement field) and that the majority of the effect will be seen on small scales adjacent to the boundary. We therefore define an edge galaxy as one that is within $10\mpcoh$ of a survey boundary and we only consider edges in the angular projection of the survey due to the low density of galaxies at the highest and lowest redshifts (as shown in Fig.~\ref{figure:nz}).

To test the impact of the mask on the recovered BAO fit values for the CMASS sample, a new mask was constructed that is cut back in angular area by $\sim20\mpcoh$ around the survey edges. Galaxies and randoms were cut using this new mask (discarding $\sim 2\%$ of each) and the displacement field was calculated using the both the full and cut regions. We refer to the masked sample as the ``cut''  sample and reconstruct it using either the overdensity of the original full survey, or only using the cut survey. We use the results from reconstruction generated from the full survey overdensity as an approximation of a survey with no edge to compare with a survey with an edge.

The BAO-scale results for both samples are given in Table~\ref{table:edges}, and are consistent suggesting that our simple method of masking the data does not alter the performance of the reconstruction algorithm for the CMASS sample. This in turn suggests that the CMASS boundary has negligible effect on the efficiency of reconstruction. As the CMASS sample has such a low edge to volume ratio, it does not provide us with a large enough percentage of edge galaxies to quantify their effect. 

In order to test the effects of edges further, we have used the CMASS mocks to artificially create a survey with a large edge-to-volume ratio. To do this, we cut the survey into 257 stripes in right ascension, $\sim0.6$ degrees across, which translates into a comoving physical separation of approximately 14$\mpcoh$ at the effective redshift of the sample. The overdensity and thus the displacement field are calculated using data spanning from one true edge of the survey up to a synthetic edge such that it is always calculated in a region covering $\geq$ half of the whole survey volume as illustrated in Fig.~\ref{fig:stripes}. The stripe of galaxies/randoms that lies on the edge of the overdensity in each instance is reconstructed using the new displacement field. Our reconstructed stripes are then concatenated to replicate a survey where the majority of galaxies ($67\%$) lie within $10\mpcoh$ of an edge. We call this ``the edge catalogue''.

On a mock by mock basis, the $\sigma_{\alpha, \mathrm{post}}$ values for the edge catalogue are larger compared to the standard reconstruction in 559 out of 600 mocks. For the remaining 41 mocks, the error is only smaller in the edge sample by an average of $\langle\Delta \sigma_{\alpha, \mathrm{post}}\rangle=0.0004$. Histograms showing the $\alpha$ and $\sigma_{\alpha}$ distributions for each sample are shown in Appendix \ref{app:hist} in Fig.~\ref{fig:edgeHIST}. Comparing the r.m.s. displacements of the edge sample with the standard sample for the first CMASS mock, the edge sample galaxies have a r.m.s. displacement of $2.9\mpcoh$ whereas the standard sample have a rms displacement of $3.6\mpcoh$. The displacements are reduced in the edge catalogue as the overdensity field beyond the boundary is not picked up and the amplitude of the displacement field drops off towards the boundary edge, where $67\%$ of the edge galaxies reside. 

The $\langle\alpha_{\mathrm{post}}\rangle$ and $\langle \sigma_{\alpha, \mathrm{post}} \rangle$ values are shown in Table~\ref{table:edges}. Although the edge sample does not do as well as the standard reconstruction, it does notably better than the non-reconstructed set of mocks. As we have constructed the edge files to represent a worst case scenario, we conclude that even surveys with a large surface area to volume ratio should benefit from reconstruction provided the galaxy density is sufficiently large, as discussed in the previous Section. Assuming a linear relation between the percentage of edge galaxies and the reduction in effectiveness of reconstruction, we can estimate the effect that a particular survey geometry (of a contiguous volume) will have. For example a survey with $20\%$ edge galaxies should expect approximately $3\%$ increase in the error on the measurement due to edge effects compared to a survey with only $2\%$ edge galaxies. For the CMASS sample, the fractional increase in the $\sigma_{\alpha, \mathrm{post}}$ Y for a specific fraction of edge galaxies X is
\begin{equation}
\mathrm{Y} = \frac{2\times 10^{-3}}{\sigma_{0}} \mathrm{X},
\end{equation}
where $\sigma_{0}$ is the error for a sample with no edges; this is $0.01116$ for the CMASS mocks in our standard reconstruction.
As the absolute value $\sigma_{0}$ is dependent on other factors, including the density and volume and redshift of sample which may not be independent of the edge results, we use this as a rough indication of the expected increase of $\sigma_{\alpha, \mathrm{post}}$ with edge fraction to show that the effect is small.

These tests have been conducted to look at edge effects on a contiguous survey, not surveys that are constructed from disjointed patches. Small holes within a survey, that are significantly smaller than the smoothing length applied, are simply equivalent to a reduction in the sample density. However, holes comparable to the smoothing scale or larger, could exclude regions important for the reconstruction as discussed in \citet{2007ApJ...664..675E}. Previous applications of reconstruction such as \citep{2012MNRAS.427.2132P} have used constrained realisations or Wiener filter methods to fill-in regions outside the survey or holes within the survey. However, it is important to realise that these methods are not providing extra information in these regions: they simply provide a plausible continuation of the density field. The efficiency of reconstruction would still be reduced near the boundaries of large holes. From the tests above we conclude that the actual effect of the boundaries is itself small for BAO-scale measurements, and this suggests that it may be unnecessary to perform a complicated extrapolation of the density field to regions where there is no data.

 \begin{table}
 \centering
  \caption{BAO scale errors recovered varying the percentage of the survey that lies along an edge.}
  \label{table:edges}
  \begin{tabular}{@{}llrrrrlrlr@{}}
  \hline
Sample & \% edge galaxies &
 $\langle\alpha_{\mathrm{post}}\rangle$    & $\langle \sigma_{\alpha, \mathrm{post}}\rangle$ & $S_{\alpha, \mathrm{post}}$ \\
 \hline
Cut   & 0   & 1.0002 & 0.0114 & 0.0113\\
         & 2   & 1.0000 & 0.0114 & 0.0114 \\
\hline
Full  & 67 & 1.0005 & 0.0125 & 0.0134\\
         & 2   &  1.0002 & 0.0112 & 0.0110\\
\hline
\end{tabular}
\end{table}
\section{Change in effectiveness with method}\label{sec:method}
\subsection{Smoothing length}  \label{subsec:smoothing}
 As discussed in Section~\ref{sec:recon}, the smoothing dictates the minimum scale of perturbations used to calculate the displacements and sets the scale on which the overdensity is measured.
\citet{PhysRevD.79.063523} noted that in theory, if the measured overdensity field were the linear matter field, and no smoothing was applied, the Zel'dovich displacements would take the data back to Lagrangian positions, and the displacements would be transferred to the random catalogue. This process would be equivalent to performing no reconstruction. 
However, working with a discrete non-linear galaxy distribution, the density field smoothed on small scales will be dominated by incoherent highly non-linear fluctuations and shot noise and we will not be correcting for the damping where the BAO signal is the strongest.
For a large smoothing scale, the algorithm will only pick up modes of the density field that are well in the in the linear regime and density perturbations in the quasi-linear regime get washed out making the algorithm less effective.
In this section we empirically measure the optimal smoothing scale.

In previous work \citep{2007ApJ...664..675E, 2012MNRAS.427.2132P,2013arXiv1312.4877A,2012arXiv1203.6594A}, a Gaussian smoothing kernel of  $R = 10 -20\mpcoh$ has been used and mildly deviating from this has been shown not to alter the results (see Appendix B of \citealt{2012arXiv1203.6594A} and \citealt{2012MNRAS.427.2132P}). Here we provide a more extensive test on how the smoothing length alters the measurements and their errors. A wide range of smoothing lengths between $5\hompc$ and $40\hompc$ on the CMASS and LOWZ mocks are considered.
 
\begin{figure*}
    \centering
    \resizebox{0.46\textwidth}{!}{\includegraphics{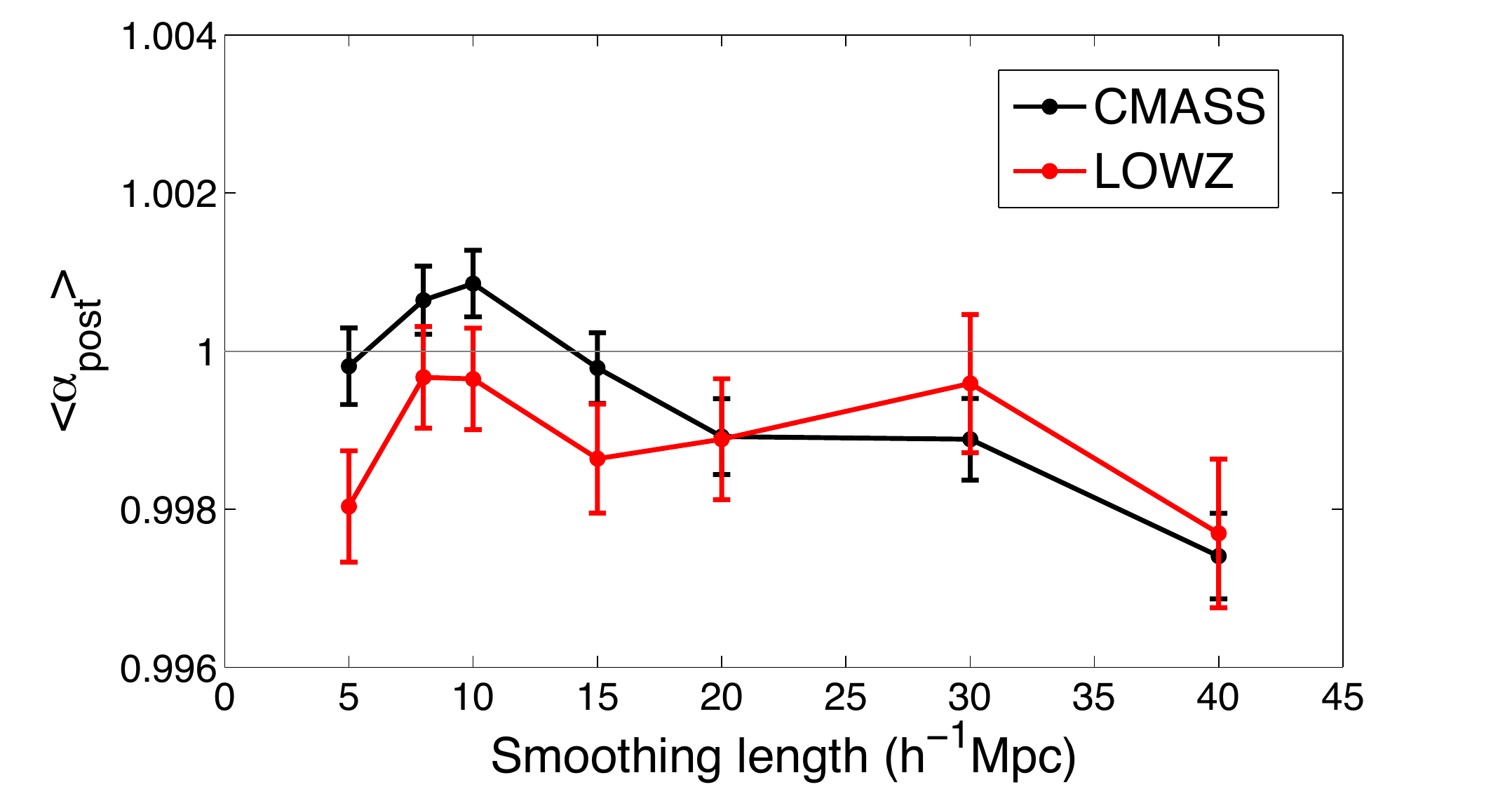}}
    \hspace{1cm}
    \resizebox{0.46\textwidth}{!}{\includegraphics{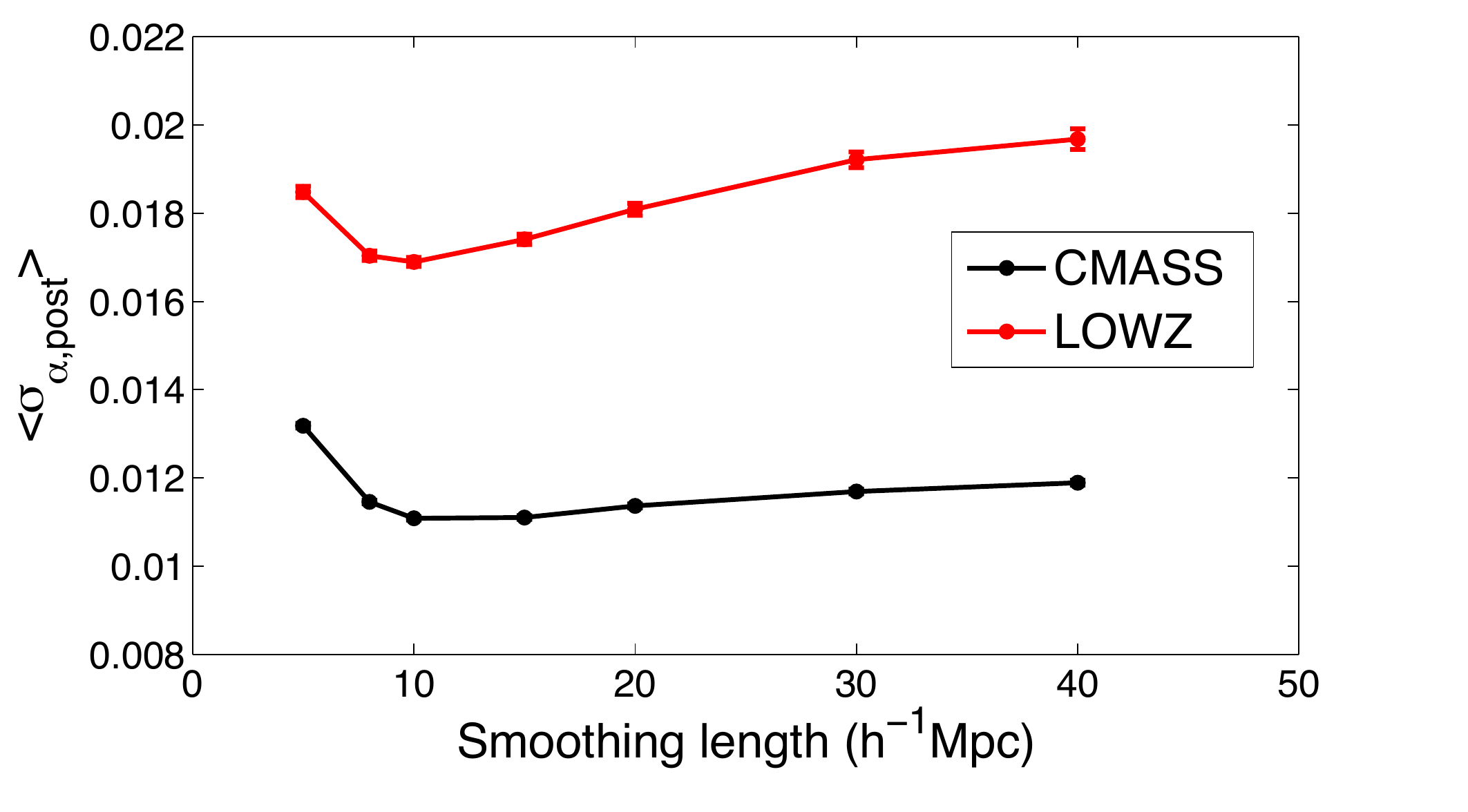}}
    \caption{The recovered $\langle\alpha_{\mathrm{post}}\rangle$ (left) and $\langle\sigma_{\alpha, \mathrm{post}}\rangle$ (right) values as a function of smoothing scale for CMASS (black line) and LOWZ (red line). The optimal smoothing scales are where the bias on $\langle\alpha\rangle$ is removed and the error $\langle\sigma\rangle$ is a minimum. The CMASS sample has an optimal smoothing scale of 15$\mpcoh$ and the LOWZ sample has an optimal smoothing scale of 10$\mpcoh$.}
  \label{fig:smooth}
\end{figure*}

Fig.~\ref{fig:smooth} shows how the smoothing scale affects $\langle\alpha\rangle$ and $\langle\sigma_{\alpha}\rangle$ recovered from the mocks.
The bias in the measurement of $\alpha$ is reduced most using the 5$\mpcoh$ and 15$\mpcoh$ for CMASS and 8$\mpcoh$ and 10$\mpcoh$ for LOWZ. For a larger smoothing scale the bias is reduced from the pre reconstruction value but the samples tend to become biased low. 
In the CMASS mocks the $\langle\sigma\rangle$ value is reduced the most with a smoothing scale of 10$\mpcoh$ and 15$\mpcoh$. In the LOWZ measurements the $\langle\sigma\rangle$ value is reduced the most with a smoothing scale of 10$\mpcoh$. When the scale is smaller than this, the algorithm quickly breaks down due to the increased non-linear and shot noise contribution to the estimate of the displacements and the error increases sharply. Conversely when the smoothing scale is increased, the result is a steady decline in the error reduction.


Below the optimal smoothing length, the reconstructed catalogues still perform better than the pre-reconstruction data. For the CMASS sample all smoothing lengths between $8\mpcoh$ and $40\mpcoh$ give an improvement on every mock and the $5\mpcoh$ smoothing kernel gives an improved result in $595$ out of the $600$ mocks. For the LOWZ sample, all smoothing lengths give an improvement in over $96\%$ of the mocks.
The average values of the best fit $\alpha$ and $\sigma_{\alpha}$ values are shown for each smoothing scale for both samples are shown in Table~\ref{table:BAO_smoothing}.
From these results we deduce that the optimal smoothing scale for CMASS is 15$\mpcoh$ and 10$\mpcoh$ for LOWZ.
  \begin{table*}
 \centering
 \begin{minipage}{130mm}
  \caption{BAO scale errors recovered for different smoothing lengths from the LOWZ and CMASS mocks.}
      \label{table:BAO_smoothing} 
  \begin{tabular}{@{}lllllc@{}}
  \hline
 Sample & Smoothing ($\mpcoh$)  &    
 $\langle\alpha_{\mathrm{post}}\rangle$    & $\langle \sigma_{\alpha, \mathrm{post}}\rangle$ & $S_{\alpha, \mathrm{post}}$ &   \% mocks with $\sigma_{\alpha,\mathrm{post}} <\sigma_{\alpha, \mathrm{pre}}$\\
 \hline
CMASS&  5  & 0.9998  & 0.0137 & 0.0118 & 99.1\%\\ 
              &  8  &  1.0006 & 0.0115 & 0.0106 & 100\%\\
	    &  10 & 1.0009  & 0.0111 & 0.0103 & 100\%\\ 
             &  15 & 0.9998  & 0.0111 & 0.0110 & 100\%\\ 
             &  20 & 0.9989  & 0.0118 & 0.0118 & 100\%\\ 
             & 30  & 0.9989 & 0.0121 & 0.0127 & 100\%\\
             &  40 & 0.9974  &  0.0124 & 0.0133 & 100\%\\ 
 \hline
LOWZ  &  5 &  0.9980 & 0.0185& 0.0172 & 96.6\%\\ 
             &  8 &  0.9997 &0.0170 &  0.0157& 99.7\%\\
             & 10 & 0.9997 & 0.0169 & 0.0157& 99.7\%\\
             & 15 & 0.9986 & 0.0174 & 0.0169 & 98.6\%\\
             & 20 & 0.9989 & 0.0181 & 0.0187 & 97.0\%\\
             & 30 & 0.9996& 0.0192 & 0.0214 & 98.3\%\\
             & 40 & 0.9977&  0.0197& 0.0231 & 98.2\%\\

\hline
\end{tabular}
\end{minipage}
\end{table*}
\subsection{Number of randoms}
\label{sec:randoms}
The random catalogue serves a dual purpose; it is compared to the galaxy density to estimate the overdensity field and it is moved in the reconstruction process where it becomes the {\em shifted} field ($\delta_{s}$). 
As it is a discrete field, it is desirable to have a large number of data points to reduce the shot noise contribution to both of these measurements. However, the reconstruction process requires a unique set of shifted randoms for each mock and as such, data storage can be a problem if these files are large. In this section we vary the number of randoms used, perform the reconstruction and compare the power spectrum fitting results.

We reduce the number of randoms in each catalogue to 10, 25 and 50 times the number of data points. As a precaution to prevent spurious correlations between mocks caused by using the same set of randoms, we randomly subsample these for each mock from the initial random catalogue of 100 times the number of data points. To prevent correlations between the displacements induced and $\delta_s$ used to calculate the 2-point statistics, we use a different base of randoms with 100 times the number of data points for each. Two sets of reconstructed catalogues are created; one using the smaller number of randoms for both fields which we name $R_{i,i}$ where $i$ is the ratio of randoms to data points in both; and one that maintains 100 times the number of randoms to calculate the overdensity but uses the smaller number of randoms in the shifted catalogue, we name these $R_{100,j}$ where $j$ is the ratio of randoms to data points in the shifted field.

\begin{figure*}
    \centering
    \resizebox{0.46\textwidth}{!}{\includegraphics{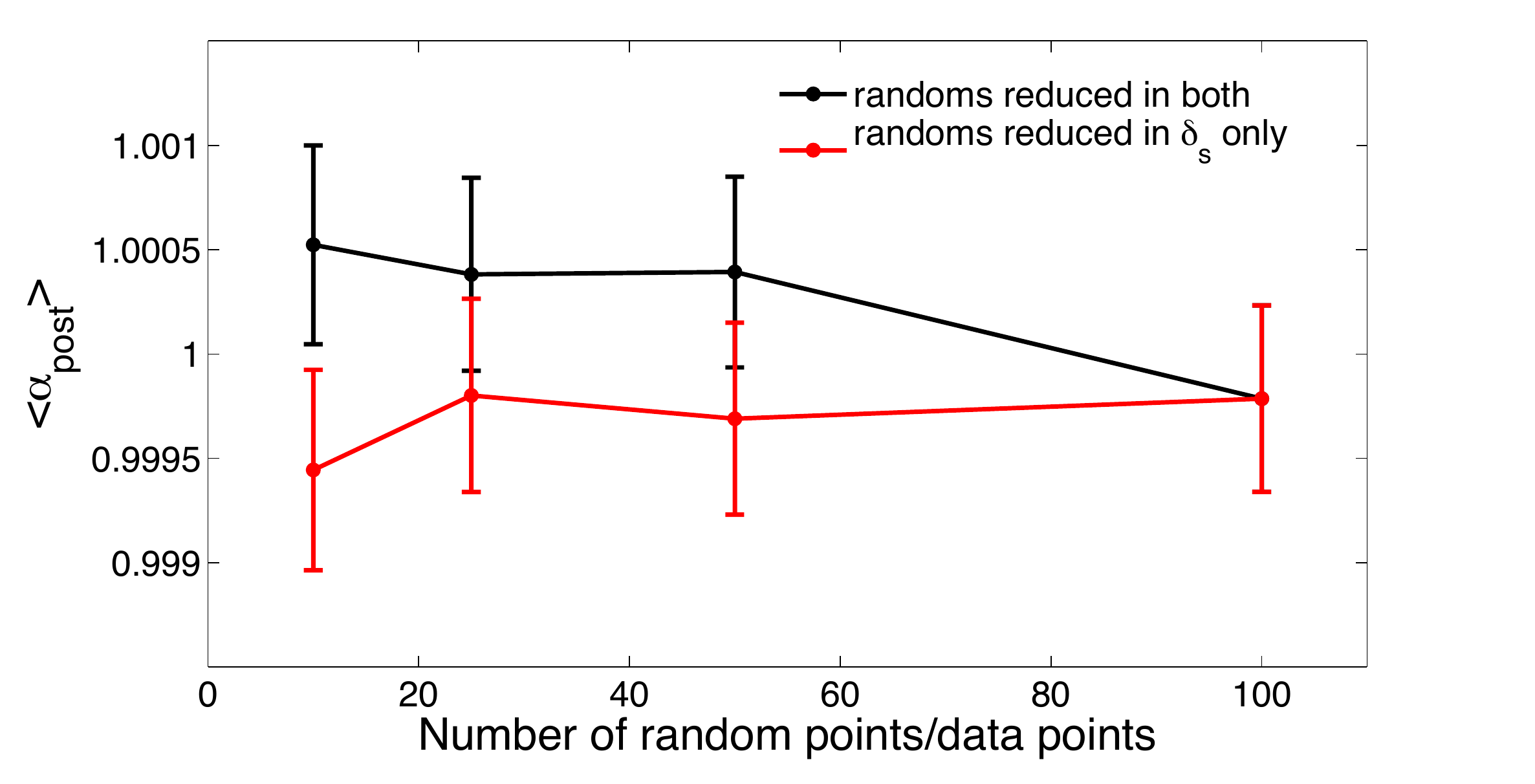}}
    \hspace{1cm}
    \resizebox{0.46\textwidth}{!}{\includegraphics{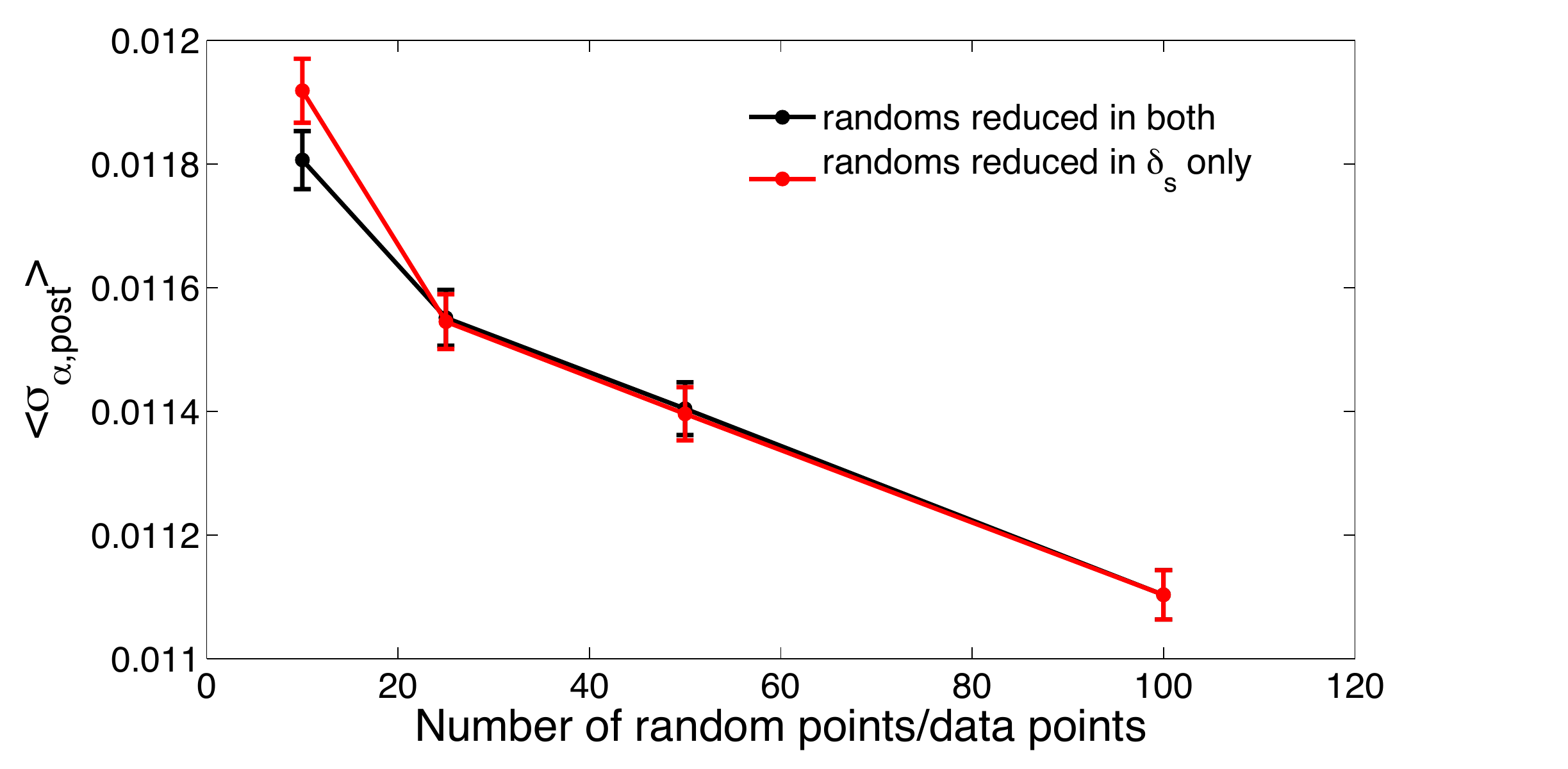}}
    \caption{The black line shows the recovered $\langle\alpha_{\mathrm{post}}\rangle$ (left) and $\langle\sigma_{\alpha, \mathrm{post}}\rangle$ (right) for CMASS catalogues reconstructed using N times the number of random points to data points (where N is the value on the x-axis) in both random catalogues. The red line shows the same recovered values for CMASS catalogues reconstructed using 100 times the number of random to data points in the overdensity calculation and N times the number of randoms to data points in the shifted random catalogue. Above N=10, both types of reconstructed catalogue show consistent measurement values. The error decrease with increasing N value suggests optimal reconstruction requires at least 25 times the number of random to data points in the overdensity calculation and as high as possible ratio of random to data points in the shifted random catalogue.}
  \label{fig:scatter_rands}
\end{figure*}

Fig.~\ref{fig:scatter_rands} shows $\langle\alpha_{\mathrm{post}}\rangle$ and $\langle\sigma_{\alpha, \mathrm{post}}\rangle$ as a function of the number of randoms for both cases. 
Both data sets have a $\langle\alpha_{\mathrm{post}}\rangle$ consistent with one for $i,j\geq25$. The $\langle\sigma_{\alpha, \mathrm{post}}\rangle$ values are consistent implying that the precision of the result is only sensitive to the number of randoms in the shifted field and increasing the number of randoms in the initial overdensity field is inconsequential as this field is smoothed. Note that the galaxy field is also smoothed, but its shot-noise is dominant and, unlike the randoms it is strongly clustered, changing the importance of the smoothing on the field. 
In the $R_{100,10}$ and $R_{10,10}$ catalogues, the $\langle\alpha_{\mathrm{post}}\rangle$ values are no longer consistent, suggesting for either random catalogue there needs to be more than 10 times the number of randoms compared to data points.

\begin{table}
 \centering
  \caption{BAO scale errors recovered for different ratios randoms to mock data for the CMASS mocks.}
  \begin{tabular}{@{}llrrrrlrlr@{}}
  \hline
Sample &
 $\langle \alpha_{\mathrm{post}}\rangle$ & $ \langle \sigma_{\alpha, \mathrm{post}}\rangle$ &  $S_{\alpha, \mathrm{post}}$\\
 \hline
$R_{10,10}$     & 0.9994 & 0.0118 & 0.0118 \\
$R_{25,25}$     & 0.9998 & 0.0116 & 0.0114\\
$R_{50,50}$    & 0.9997 & 0.0114 & 0.0113\\
$R_{100,10}$  & 1.0005 & 0.0119 & 0.0117\\
$R_{100,25}$   & 1.0004 & 0.0115 & 0.0113\\
$R_{100,50}$   & 1.0004 & 0.0114 & 0.0112\\
$R_{100,100}$ & 0.9998 & 0.0111 & 0.0110\\
\hline
\end{tabular}
\end{table}
\subsection{Finite difference method}
\label{sec:FD}
There are a number of options for finding solutions to Eq.~(\ref{eq:maineq}), including methods based in Fourier space or in configuration space as used by \citet{2012MNRAS.427.2132P}. 
To check that the approximations used in the configuration space method of \citet{2012MNRAS.427.2132P} give the same solution as our Fourier based method, we have implemented both.
The configuration space method solves for the potential as defined in Eq.~(\ref{eq:potential}) and the equation we want to solve is Eq.~\ref{eq:maineq} rewritten in terms of the potential.
\begin{equation}\label{eq:mainpot}
\nabla^2 \phi +\frac{ f}{b} \nabla\cdot \left(\nabla\phi_r\right) \mathbf{\hat{r}} = \frac{-\delta_g}{b}.
\end{equation}
We solve this on a grid using finite differences to approximate the derivatives.
The potential at each grid point is expressed as a function of the potential at the surrounding grid points. The Laplacian of the potential at a grid point can be approximated as a function of the potential at the 6 nearest grid points and the central point. 
\begin{equation}
\nabla^2 \phi _{000} \approx \frac{1}{g^2} \left[ \sum_A \phi_{ijk}  -6\phi_{000}\right],
\end{equation}
where the sum over $A$ is the sum over the 6 adjacent grid points and $g$ is the spacing between grid points.
The second part of Eq.~\ref{eq:mainpot} can be written as
\begin{equation}
\frac{f}{b} \nabla\cdot \left(\nabla\phi_r\right) \mathbf{\hat{r}} = \frac{f}{b}\left(\mathbf{ \hat{r} \cdot \nabla \left(\nabla \phi_r \right)} + \mathbf{ \nabla} \phi_r \left(\mathbf{\nabla \cdot \hat{r}}\right) \right),
\end{equation}
which can be approximated as 
\begin{equation}
-2\frac{f}{b}\frac{\phi_{000}}{g^2} + \sum_B \frac{f}{b}\left(\frac{x_i^2}{g^2r^2} \pm  \frac{x_i}{g r^2}\right) \phi_{A} + \sum_C (-1)^p \frac{f}{b} \frac{x_i x_j }{2 g^2 r^2}\phi_B
\end{equation}
where $B$ is the set of points $ijk$ such that 2 of the indices are zero and the other is $\pm 1$.  $x_i$ the cartesian position of the non-zero index and $r$ is the distance to the central grid point. $C$ is the set of points where two of the indices are $\pm 1$ and the third is zero. When the two indices are the same, $p=0$ , when they are different $p=1$. $x_i$ and $x_j$ are the cartesian positions of the non-zero indices. 

This can be arranged as a linear system of equations such that $\mathbf{A \phi} = \mathbf{\delta}$, where $\mathbf{A}$ is a matrix describing the dependence of the potential on its surroundings. The $\delta$ that we input here is the same smoothed overdensity field as we use in the Fourier method. We solve for the potential using the GMRES in the PETSc package \citep{petsc-web-page,petsc-user-ref} as in \citet{2012MNRAS.427.2132P}.  Finite differences are used again to calculate the displacements at each grid point from the potential.

In Fourier space we solve directly for the displacement field using Fast Fourier Transforms in the FFTW package \citep{FFTW05}. 
We want to solve for $\mathbf\Psi$ in Eq.~\ref{eq:maineq} and we outline the steps in the process.
Assuming $\mathbf{\Psi}$ is irrotational then the two vector fields on the left of the equation can be expressed as gradients of scalar fields, so let
\begin{equation}
\mathbf{\Psi} = \nabla \phi,
\end{equation}
\begin{equation}
\frac{f}{b}\left(\mathbf{\Psi\cdot \hat{r}}\right)\mathbf{\hat{r}} = \nabla \gamma.
\end{equation}
Thus, Fourier transforming and carrying out the double derivatives results in
\begin{equation}
\phi\left(\mathbf{k}\right) + \gamma\left(\mathbf{k}\right) = \frac{\delta \left(\mathbf{k}\right)}{k^2 b},
\end{equation}
and so
\begin{equation}
\nabla \left( \phi\left(\mathbf{k}\right) + \gamma\left(\mathbf{k}\right)\right) = - \frac{\mathbf{i k}\delta\left(\mathbf{k}\right)}{k^2 b}.
\end{equation}
and finally
\begin{equation}
 \mathbf{\Psi} + \frac{f}{b} \left( \mathbf{\Psi \cdot \hat{r}} \right) \mathbf{\hat{r}} ={\rm IFFT} \left[- \frac{\mathbf{i k }\delta\left(\mathbf{k}\right)}{k^2 b}\right]
\end{equation}
In cartesian coordinates this gives three equations that can be solved simultaneously to get $\Psi_x$, $\Psi_y$ and $\Psi_z$. IFFT indicates the inverse Fast Fourier Transform.
 
The accuracy of the discrete Fourier transform is dependent on the sampling rate of the data, where a signal with frequency above the Nyquist limit will not be recovered. As our smoothing length is larger than our grid size we are not concerned about the loss of information at these frequencies.

\begin{figure*}
    \centering
    \resizebox{0.46\textwidth}{!}{\includegraphics{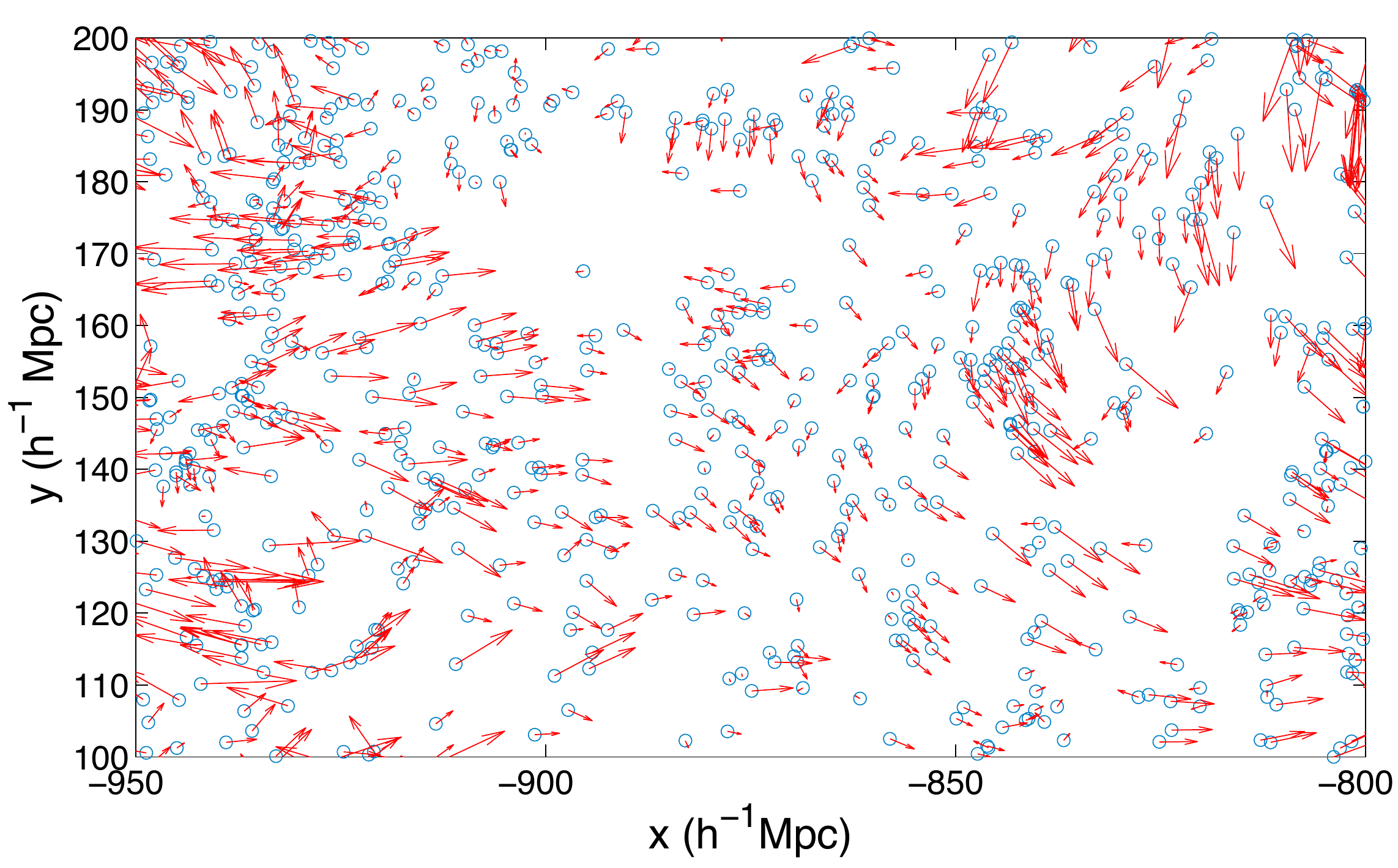}}
    \hspace{1cm}
    \resizebox{0.46\textwidth}{!}{\includegraphics{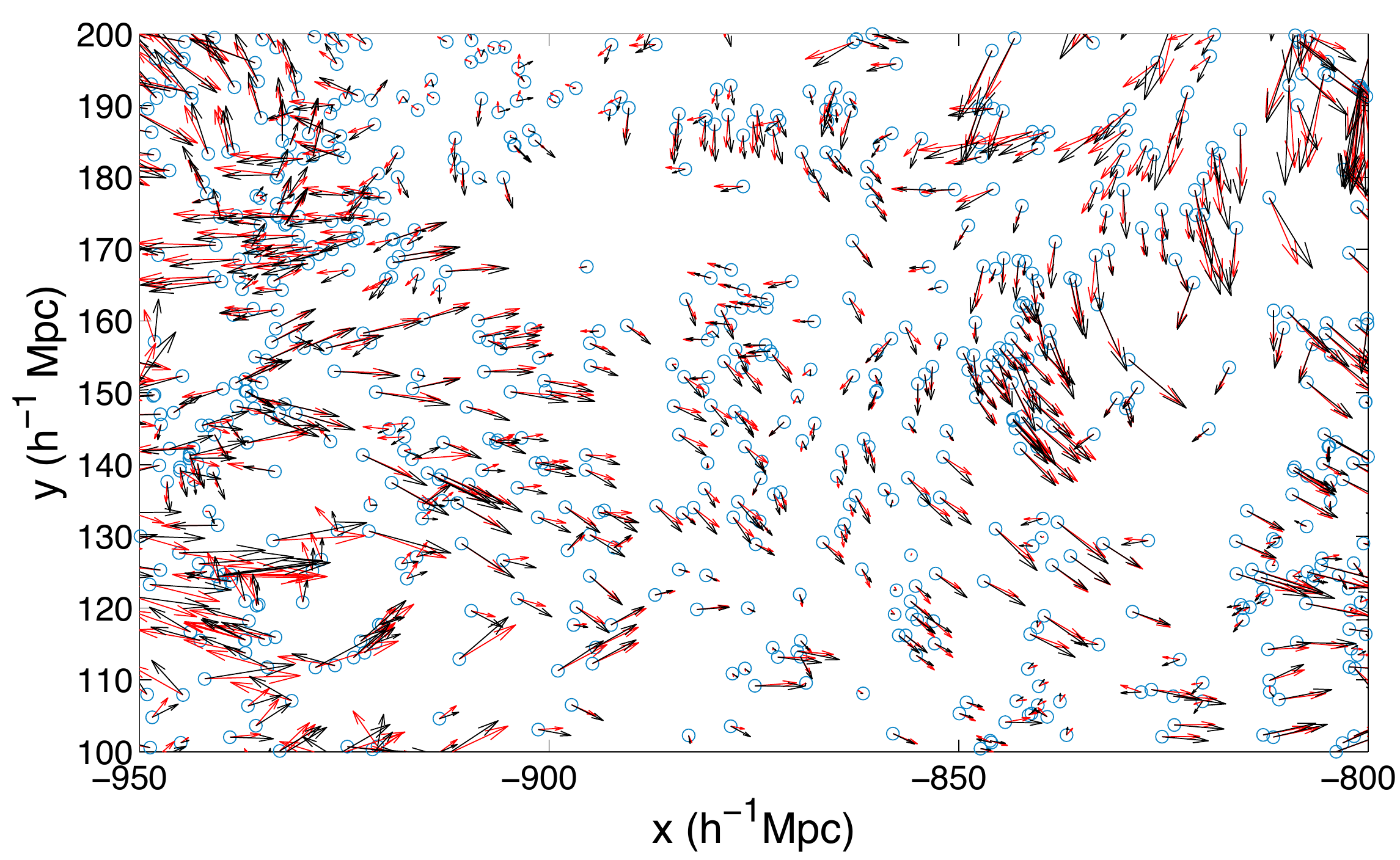}}
    \caption{Left, Lagrangian displacement field projected in 2D from finite difference method (red), the initial galaxy positions are shown with the open circles. Right, the same as on the left but with the displacement vectors from the Fourier method over plotted (black).}
  \label{fig:FD_vs_Fourier}
\end{figure*}

\begin{figure}
  \includegraphics[width=\linewidth]{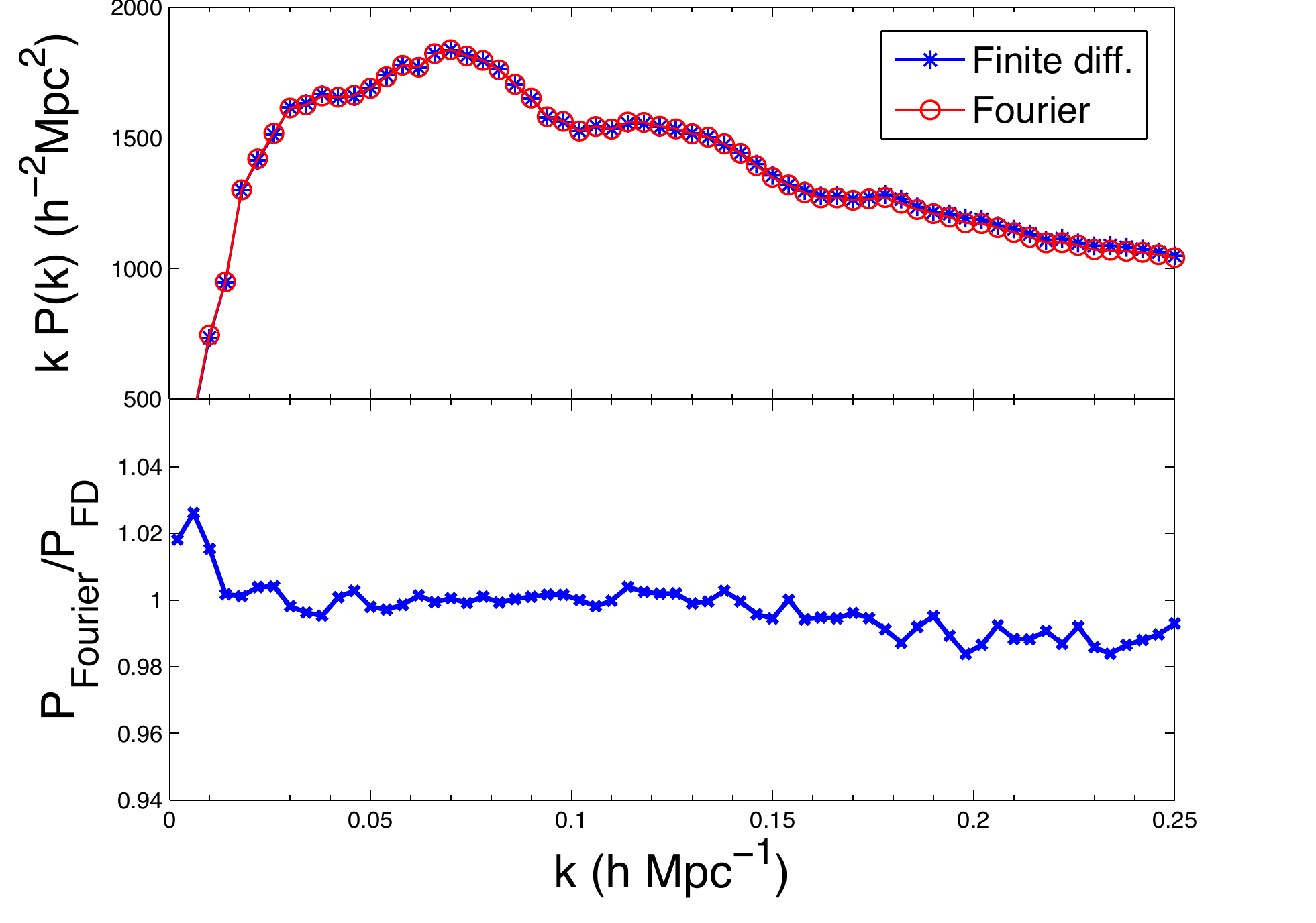}
\caption{The top panel shows a comparison of average power spectra of the first 10 LOWZ mocks reconstructed using the finite difference method (open circles) and the Fourier method (crosses). The bottom panel shows the ratio between the two set of power spectra.}
  \label{fig:FD_power}
\end{figure}

Implementing both codes, we show the comparison of displacement vectors recovered for individual galaxies for the first LOWZ mock catalogue. Fig.~\ref{fig:FD_vs_Fourier}
shows the displacement vectors projected in 2D from a slice through the survey, on the left hand side, the black vectors are from the Fourier method only, on the right hand side, the red vectors are from the finite difference method are plotted on top and the open circles are the original galaxy positions. This patch is a good representation of other regions of the survey inspected. The two vector fields are well aligned with only small differences that can be expected from using approximate methods. 
Although the amplitudes and directions of the displacements look similar for each method, this does not automatically imply that the statistical interpretation of the catalogues produced by both methods will be the same. To check that both methods will deliver the same statistical results we reconstruct the first 10 LOWZ mocks using the finite difference method and compare their power spectra to the first 10 LOWZ mocks reconstructed using our standard Fourier procedure. The average power spectra are shown in Fig.~\ref{fig:FD_power} (top) and their ratio (bottom). The ratio of power spectra show that both methods are in good agreement with deviations on small scales as expected. 

\subsection{RSD removal}\label{sec:RSD}
  \begin{table}
 \centering
  \caption{CMASS,  $\langle\sigma_{\alpha}\rangle$ with/without RSDs removed during reconstruction}
  \label{tab:RSD}
  \begin{tabular}{@{}llrrrrlrlr@{}}
  \hline
Type  &    
 $\langle\alpha\rangle$    & $\langle \sigma_{\alpha, \mathrm{post}}\rangle$ & $S_{\alpha, \mathrm{post}}$\\
 \hline
 With RSDs removed  & 1.0009  & 0.0111 & 0.0103\\ 
 Without RSDs removed & 1.0006 & 0.0112 & 0.0108\\
\hline
\end{tabular}
\end{table}
The redshift space position of a galaxy is a combined measurement of the velocity field and the real space density field. Thus the clustering along the line-of-sight is enhanced, and contains more information than across the line-of-sight. Note that there is a subtlety here - if we simply take a measured field and multiply it by a factor, we do not change the information content. What is happening in redshift space is that we are increasing the clustering strength of the underlying field but not changing the shot noise, and thus the information is increased as is the effective volume (as given in Eq.~\ref{eq:Veff}). 

However, when we remove the linear RSDs from the density field using Eq.~\ref{eq:RSD} we infer the displacement field from the redshift-space data, and thus we are not decoupling the two signals or adding/subtracting any new information. Therefore removing the redshift space distortions in this manner should not affect the signal-to-noise, but does reduce the amplitude of the power spectrum.
 by the Kaiser boost factor (which is input into the algorithm) as previously shown in Fig.~\ref{fig:power_pre_post}. 
\begin{figure}
  \includegraphics[width=\linewidth]{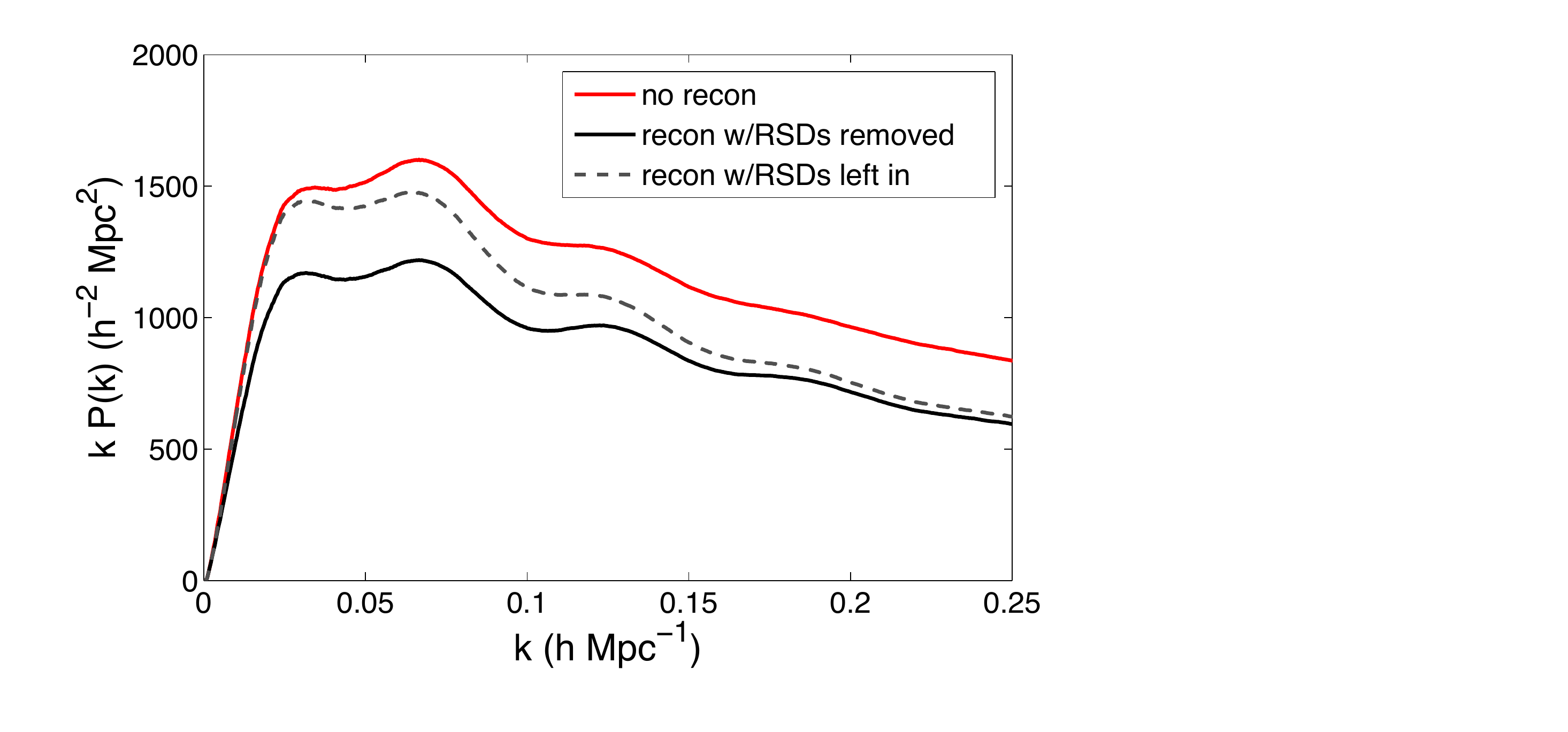}
\caption{Comparison of average power spectra from CMASS mocks with no reconstruction, reconstruction that removes RSDs and reconstruction that leaves the RSDs in the galaxy distribution.}
  \label{fig:RSDpower}
\end{figure}

We run the reconstruction code leaving the RSDs in the galaxy field and compare the $\langle\alpha_{\mathrm{post}}\rangle$ and $\langle\sigma_{\alpha, \mathrm{post}}\rangle$ values to those with the RSDs removed.
The results are shown in Table \ref{tab:RSD} where the $\langle\alpha_{\mathrm{post}}\rangle$ values and $\langle\sigma_{\alpha, \mathrm{post}}\rangle$ values are consistent. If we could measure the velocity field directly, we expect that removing the RSDs should decrease the signal to noise of the measurement. Note that by removing the RSD and changing the amplitude of the power spectrum as a function of angle to the line of sight, we are altering the relative contribution of modes to the monopole, and consequently the cosmological meaning of the BAO measurement made. 
The average power spectra are shown in Fig.~\ref{fig:RSDpower}, and compared to the pre reconstruction and standard reconstruction power spectra. 

\section{Conclusions}\label{sec:conclusion}
This paper presents the results of tests designed to optimise the efficiency of reconstruction when calculating the BAO scale from the spherically averaged power spectrum, via input parameters of the algorithm and external influences of survey design.

In all of our tests, the algorithm leads to an improvement in our ability to measure the BAO signal compared to the non-reconstructed sample and the procedure in general is found to be very robust.
However, obviously, we want to ensure reconstruction is running at maximum efficiency to extract the most precise measurements possible.

\subsection{Algorithm}
We have tested the algorithm to extract the optimal smoothing scale, determine the consequence of shot noise in the random catalogues and look for inconsistencies in the corrective bulk-flow displacements due to the method used to estimate them.

Smoothing the overdensity prior to calculating the displacement field ensures that displacements are sourced from density regions responsible for the bulk flows which cause the strongest degradation of the linear BAO signal. The Gaussian smoothing width is a free parameter in our code, and so we test  a wide range of smoothing scales. If the smoothing width is too large we only decouple modes of the density field that are already in the linear regime and suppress useful overdensity information. Conversely if the smoothing scale is too small, we decouple modes on scales smaller than the BAO signal. 
In the higher redshift sample the $\langle\alpha_{\mathrm{post}}\rangle$ becomes increasingly biased with a smoothing length greater than $15\mpcoh$. The $\langle\sigma\rangle$ values show an optimal smoothing length of between 10-15$\mpcoh$ for the higher redshift sample and 10$\mpcoh$ for the low redshift sample. In \citet{2012JCAP...10..006T}, they propose an iterative scheme to extract the particle displacements where the optimal smoothing length is calculated directly from the overdensity field at each step. We have not tested such a scheme here.

One of the practical concerns of implementing this reconstruction process is the storage of large random catalogues. There are two random catalogues used in the reconstruction process, one to set up the over density field and another that is shifted as part of the reconstruction process and combined with the reconstructed mock data to calculate the 2-point statistics.
The density fields of the mock and random catalogues are smoothed prior to calculating the overdensity. Thus increasing the number of randoms in the first catalogue does not improve the efficiency of the reconstruction algorithm provided that there 25 times plus the number of randoms to mock galaxies. However, to prevent correlations between mocks within the same sample, it is recommended that this random catalogue is different for each separate mock. 
However, the second random catalogue (which becomes the shifted random catalogue), is not smoothed. In order to reduce the shot noise in the power spectrum measurements this catalogue requires as many data points as possible. Unfortunately, each reconstruction instance produces a unique shifted random catalogue, hence storage of data may be problematic. 
Alternative solutions may be to incorporate the reconstruction into the 2-point statistic measurements calculating the random catalogues \mydoubleq{on the fly}. 

We have shown that this reconstruction algorithm generates the same displacement fields whether using finite difference approximations in configuration space or Fourier based methods.
Furthermore, we have shown that the method of inferring the RSDs from the same Lagrangian displacement field used in the reconstruction process, does not change the signal to noise of the reconstructed catalogues, but only reduces the amplitude of the clustering on large scales via our input values of bias and the growth function.

To summarise, we recommend using a smoothing length of between 10-15 $\mpcoh$, and as many points in the reconstructed shifted random catalogues as storage will permit. We find no difference between Fourier and configuration space methods of estimating the displacement field and show that the method of removing RSDs used does not alter the signal to noise of the measured BAO signal.

\subsection{Survey design}
We have examined the efficiency of reconstruction versus external factors of the survey; galaxy density, survey volume and edge to volume ratio which will provide repercussions for future survey design.

The density of the survey has the greatest impact on the reconstruction algorithm within the bounds of our test parameters. This should come as no surprise as the survey contains the information used in the reconstruction process. For a given survey density we can predict how well reconstruction should perform.
We test mock catalogues at two redshifts, z=0.32 and z=0.57. Reconstruction removes the expected bias in the measurements at all densities for the higher redshift samples. For the lower redshift samples the detection of the bias is only apparent at $V_{eff}>0.4\gpcohV$. Reconstruction removes the bias in those cases. 

Initially, the low redshift sample has a higher error pre reconstruction as a function of effective volume due to a greater non-linear component of its density field. The error on the measurement for both samples is reduced both pre and post reconstruction as $V_{eff}$ is increased. 
To separate the improvement due to increased volume from the improvement due to reconstruction we look at the percentage reduction in error as a function of the average survey density. Both samples show a strong trend of increasing efficiency of reconstruction with increased density with no indication of asymptoting to an optimal density. We perform a linear fit to the data which suggests that for reconstruction to reduce the error on the measurement to half of its pre reconstruction value requires a survey density of $\approx 4\times 10^{-4}\hompcV$ on average. 

For surveys with large edge to volume ratios, we have provided an estimate of the reduction in precision expected due to edge effects. The effects are very small; for our worst case sample containing $67\%$ of galaxies less than $10\mpcoh$ from a survey boundary, $\sigma_{\alpha, {\rm post}}$ is only increased by $12\%$. We expect for surveys with less than $5\%$ of galaxies within $10\mpcoh$ of a boundary, the increase in $\sigma_{\alpha, {\rm post}}$ due to edge effects will be negligible. Linearly extrapolating the results, the increase in error is only $3\%$ for every extra $20\%$ of edge galaxies. 

To summarise, we suggest that the strong density dependence on efficiency of the algorithm will change the optimal balance between density and volume, and should be considered by future surveys. A higher density over larger volumes is desirable to optimise the post-reconstruction BAO measurement errors. For a survey with a contiguous volume, we find that a high edge to volume ratio does not have a big impact on the efficiency of reconstruction.

\null

In conclusion, we have shown that reconstruction algorithm is a robust method of improving monopole measurements of the BAO scale. Although it is robust, there are ways of optimising the efficiency of the algorithm with regards to the methodology including smoothing length and the number of random data points to use. We have made predictions of the expected improvement from reconstruction from our tests for survey density and the volume to edge ratio.

We believe that our paper provides a step forwards in the practical implementation of reconstruction for surveys, and will also aid in the design of future surveys where the trade between sample density and reconstruction efficiency can now be predicted. However, there are many further tests to perform, including the analysis of reconstruction with regards to the anisotropic clustering measurements, considering cosmological dependencies, and further statistical methods for calculating the density field. These are left for future work.

\section*{Acknowledgments}
Many thanks to Nikhil Padmanabhan for useful correspondence. Thanks to Gary Burton, the HPC support and SCIAMA administrator for accommodating 270 reconstruction runs for each of the 600 CMASS mocks and 13 reconstruction runs for the 1000 LOWZ mocks. 
WJP acknowledges support from the UK STFC through the consolidated grant ST/K0090X/1, and from the European Research Council through grants MDEPUGS and Darksurvey.
Funding for SDSS-III has been provided by the Alfred P. Sloan Foundation, the Participating Institutions, the National Science Foundation, and the U.S. Department of Energy Office of Science. The SDSS-III web site is http://www.sdss3.org/.

SDSS-III is managed by the Astrophysical Research Consortium for the Participating Institutions of the SDSS-III Collaboration including the University of Arizona, the Brazilian Participation Group, Brookhaven National Laboratory, Carnegie Mellon University, University of Florida, the French Participation Group, the German Participation Group, Harvard University, the Instituto de Astrofisica de Canarias, the Michigan State/Notre Dame/JINA Participation Group, Johns Hopkins University, Lawrence Berkeley National Laboratory, Max Planck Institute for Astrophysics, Max Planck Institute for Extraterrestrial Physics, New Mexico State University, New York University, Ohio State University, Pennsylvania State University, University of Portsmouth, Princeton University, the Spanish Participation Group, University of Tokyo, University of Utah, Vanderbilt University, University of Virginia, University of Washington, and Yale University.
\bibliography{Efficient_recon_linear_BAO_v1}
\appendix
\section{Erratum:}
Although ${\bf \Psi}$ is irrotational, $ ({\bf \Psi \cdot \hat{r}}) {\bf \hat{r}}$ is not, and therefore Eq's. 30 - 33 are not exact. Performing a Helmholz decomposition 
\begin{equation}\label{eq:two}
({\bf \Psi\cdot \hat{r}}){\bf \hat{r}} =  \nabla A + \nabla \times {\bf B},
\end{equation}
where A is a scalar potential field and ${\bf B}$ is a vector potential field, we see that Eq. 33 only picks up the scalar potential field component. In a plane parallel approximation, ${\bf \hat{r}} \rightarrow {\bf \hat{x}} $, the non-zero $\nabla \times {\bf B}$ component has terms
\begin{equation*}
(\nabla \times {\bf B})_{{\bf x}} =  \frac{1}{k^2} \left( k_y^2 + k_z^2\right) \Psi_x {\bf \hat{x}},
\end{equation*}
\begin{equation*}
(\nabla \times {\bf B})_{{\bf y}} = - \frac{k_y k_x }{k^2} \Psi_x {\bf \hat{y}},
\end{equation*}
\begin{equation*}
(\nabla \times {\bf B})_{{\bf z}} = - \frac{k_z k_x }{k^2} \Psi_x {\bf \hat{z}},
\end{equation*}
which are missed, and thus Eq. 33 should only be considered an approximation.

However, the correction does account for the irrotational component of $ ({\bf \Psi \cdot \hat{r}}) {\bf \hat{r}}$ and thus improves the initial measurement of the ${\bf \Psi}$ field from the data and is therefore better than ignoring RSD.

The BOSS galaxy samples used in this work are highly biased, $b\approx 2$, compared with the growth function $f\approx 0.7$, which makes the correction to the displacement field calculation from the redshift space density field small. Thus the derived displacement field is only weakly dependent on the RSD. This is supported by the empirical comparison of Fourier and finite difference methods in Section 7.3 where very similar isotropic results were presented. Thus we believe that the results and primary conclusions from our work remain valid, although we caution against using this Fourier method to compute anisotropic measurements without further testing and/or development.

\section{Histograms}
\label{app:hist}
\begin{figure*}
\minipage{1\textwidth}
    \centering
    \resizebox{0.46\textwidth}{!}{\includegraphics{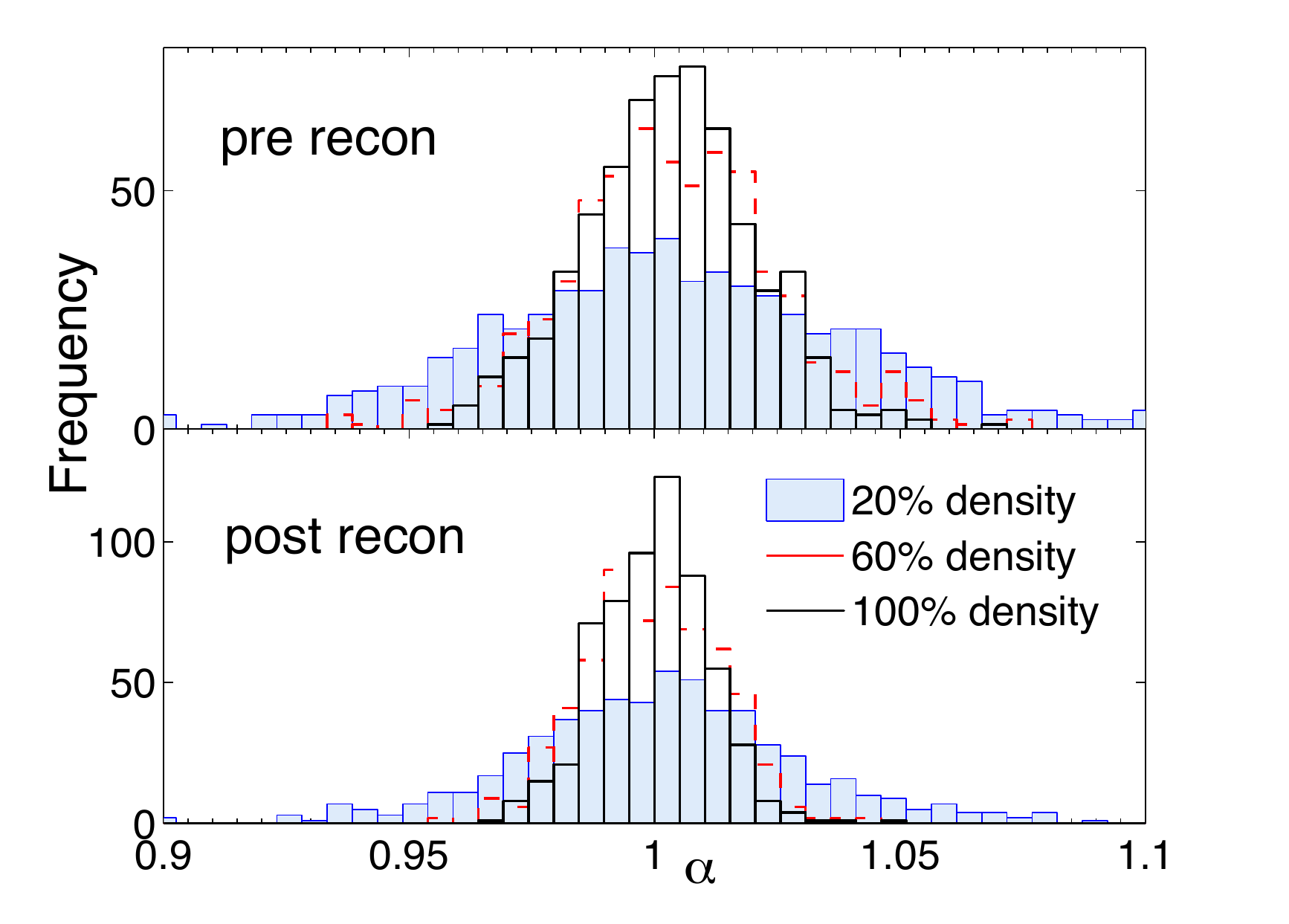}}
    \hspace{1cm}
    \resizebox{0.46\textwidth}{!}{\includegraphics{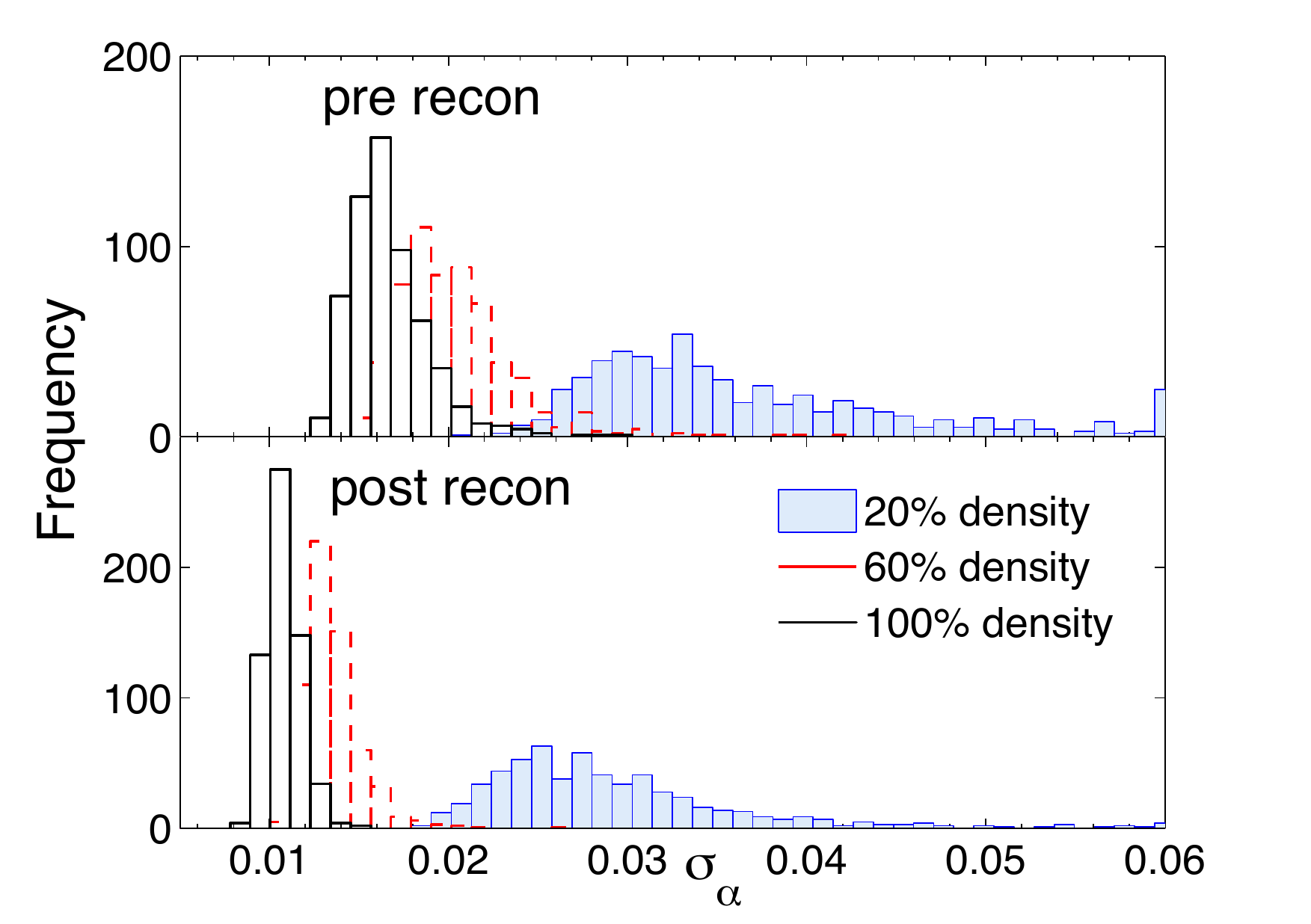}}
\endminipage\hfill
	\caption{ Distribution in $\alpha$ and $\sigma_{\alpha}$ for the CMASS mocks at different densities pre and post reconstruction.}
	\label{fig:CMASSdensityHIST}
\end{figure*}

\begin{figure*}
\minipage{1\textwidth}
    \centering
    \resizebox{0.46\textwidth}{!}{\includegraphics{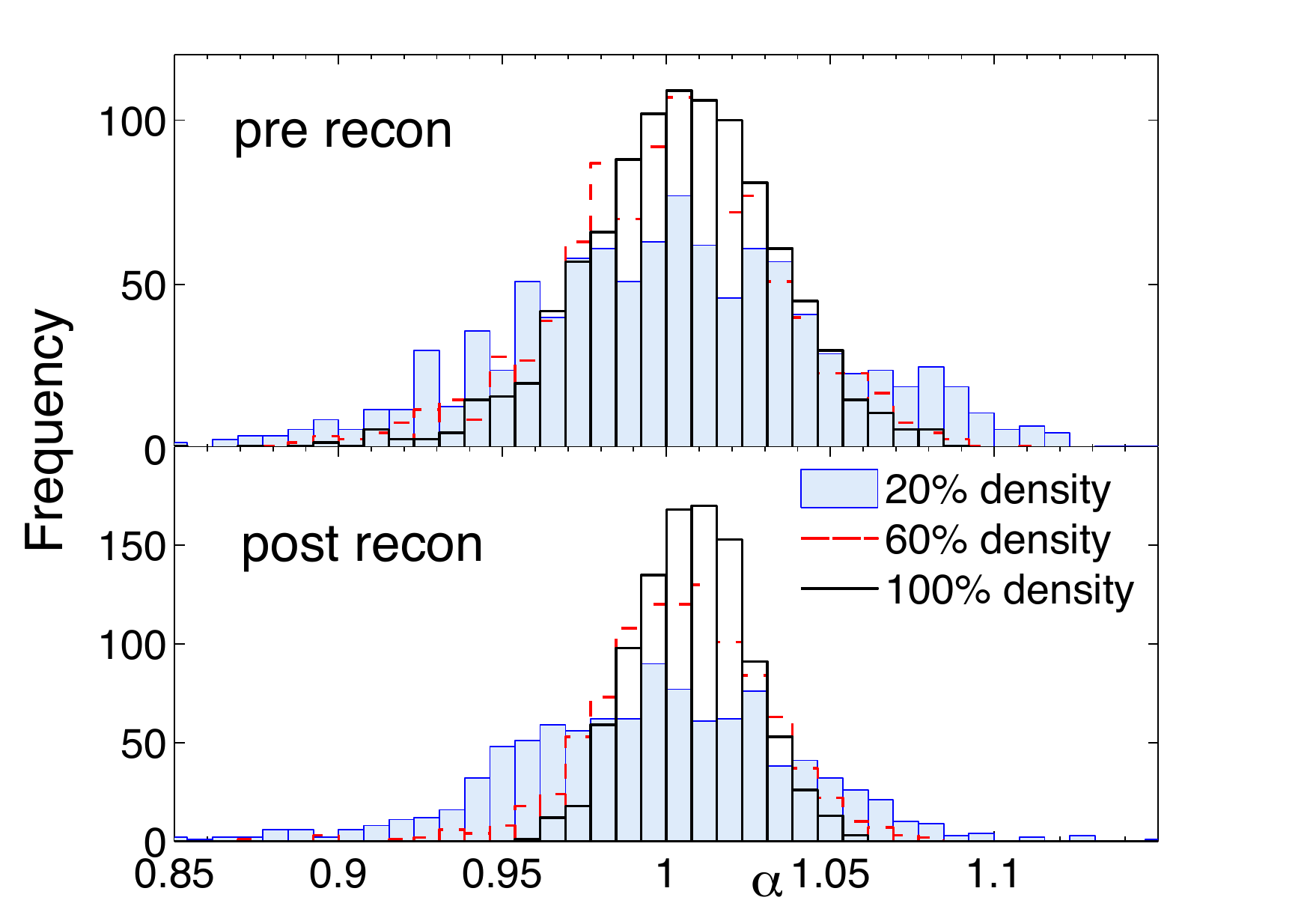}}
    \hspace{1cm}
    \resizebox{0.46\textwidth}{!}{\includegraphics{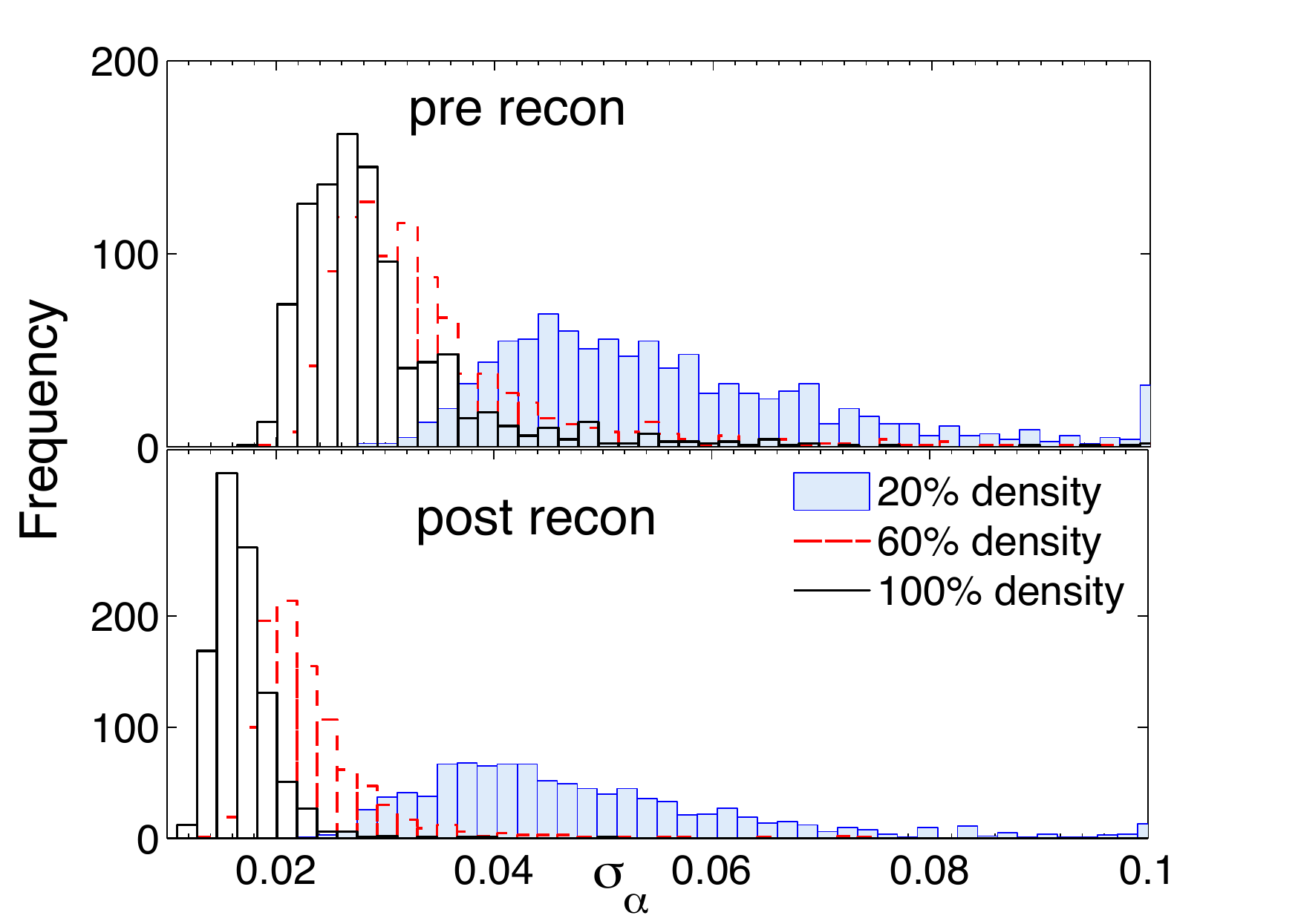}}
\endminipage\hfill
	\caption{ Distribution in $\alpha$ and $\sigma_{\alpha}$ for the LOWZ mocks at different densities pre and post reconstruction.}
	\label{fig:LOWZdensityHIST}
\end{figure*}

\begin{figure*}
    \centering
    \resizebox{0.46\textwidth}{!}{\includegraphics{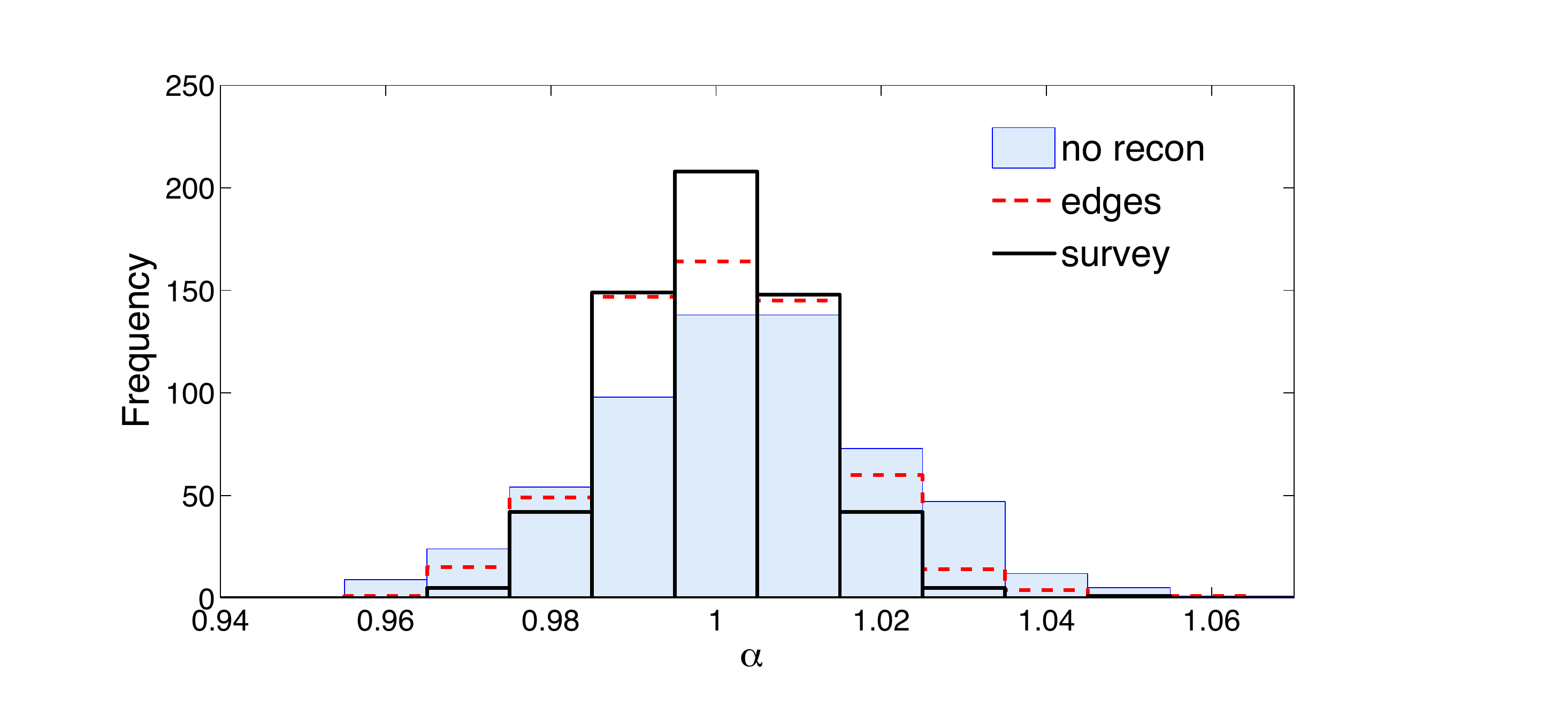}}
    \hspace{1cm}
    \resizebox{0.46\textwidth}{!}{\includegraphics{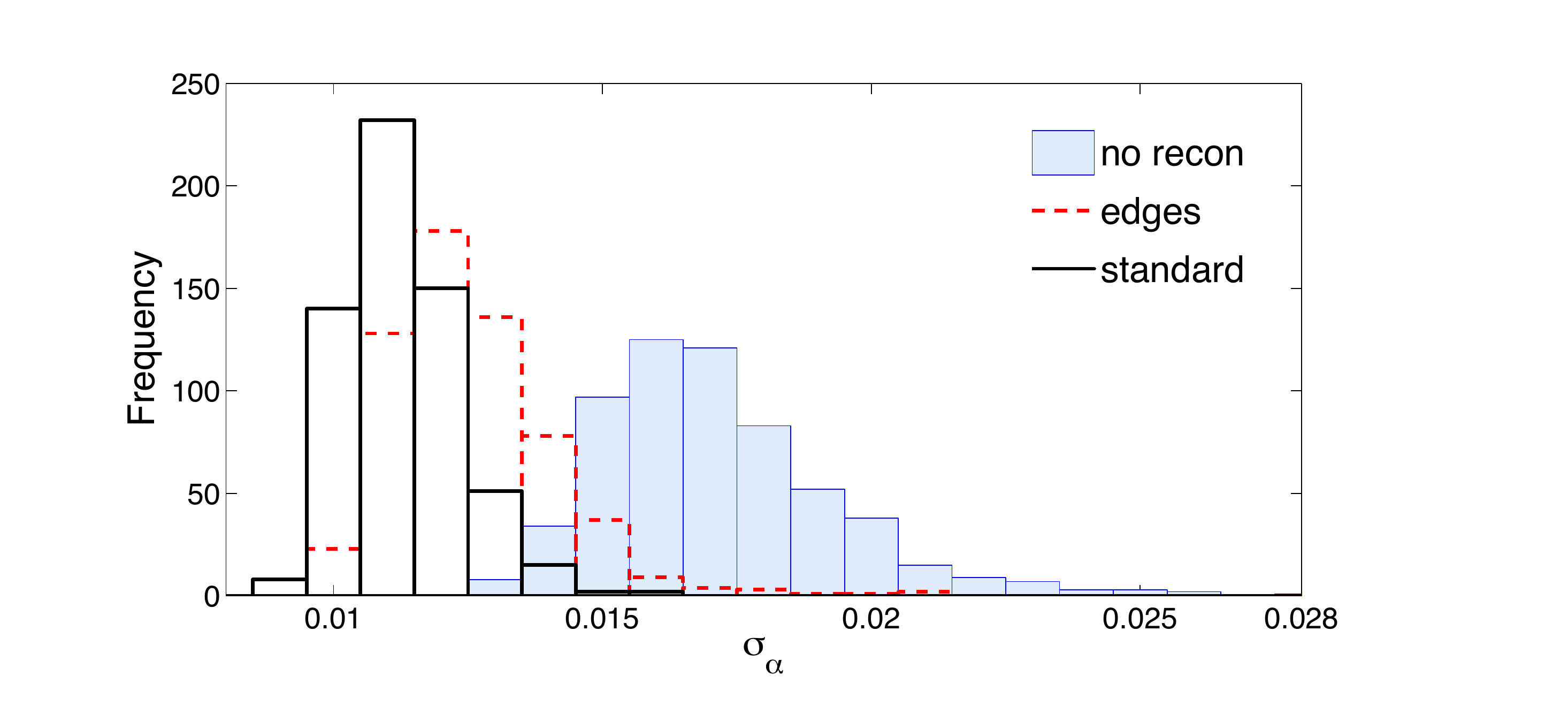}}
	\caption{ Distribution in $\alpha$ and $\sigma_{\alpha}$ for the CMASS edges sample compared to the standard reconstructed and non-reconstructed samples.}
	\label{fig:edgeHIST}
\end{figure*}

\begin{figure*}
\minipage{1\textwidth}
    \centering
    \resizebox{0.46\textwidth}{!}{\includegraphics{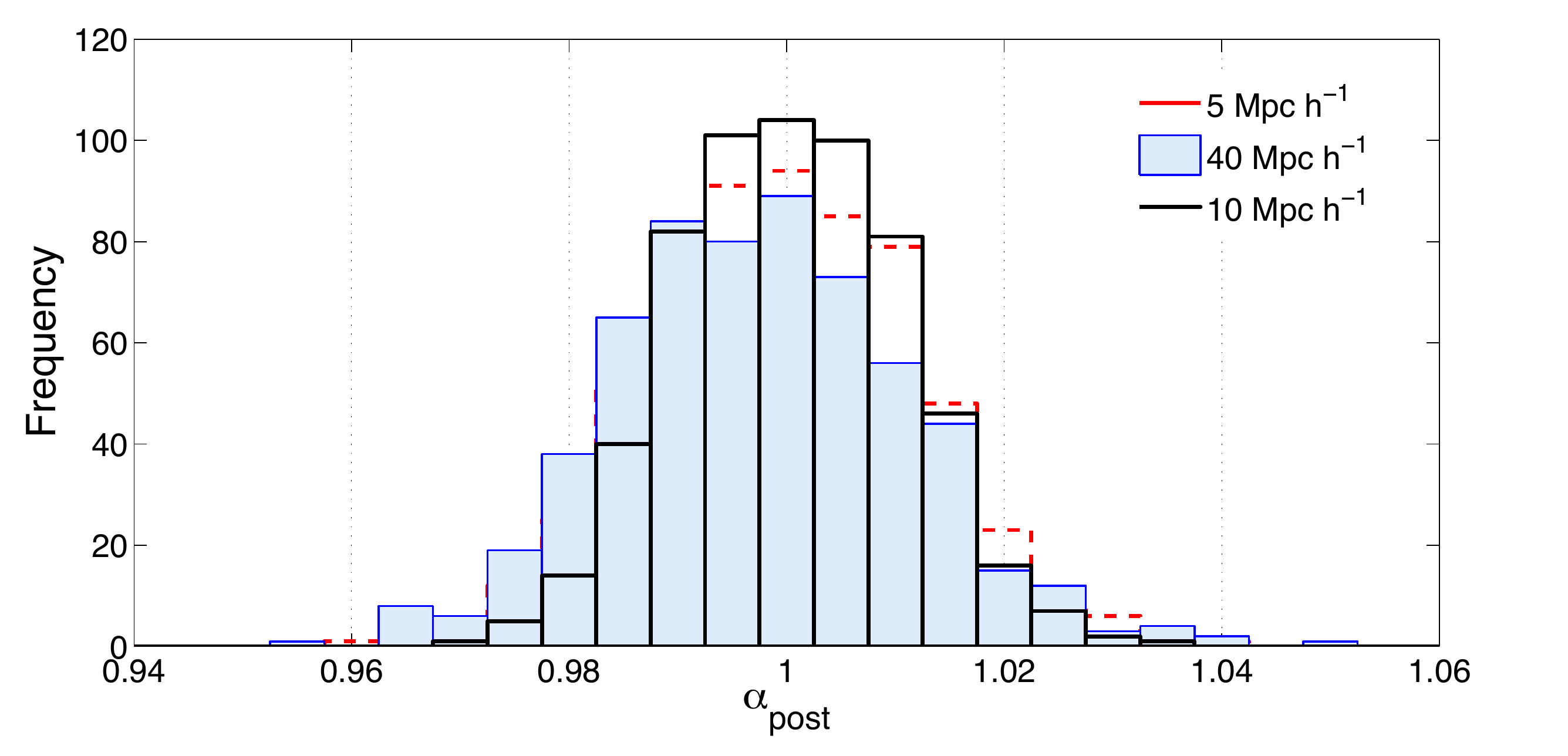}}
    \hspace{1cm}
    \resizebox{0.46\textwidth}{!}{\includegraphics{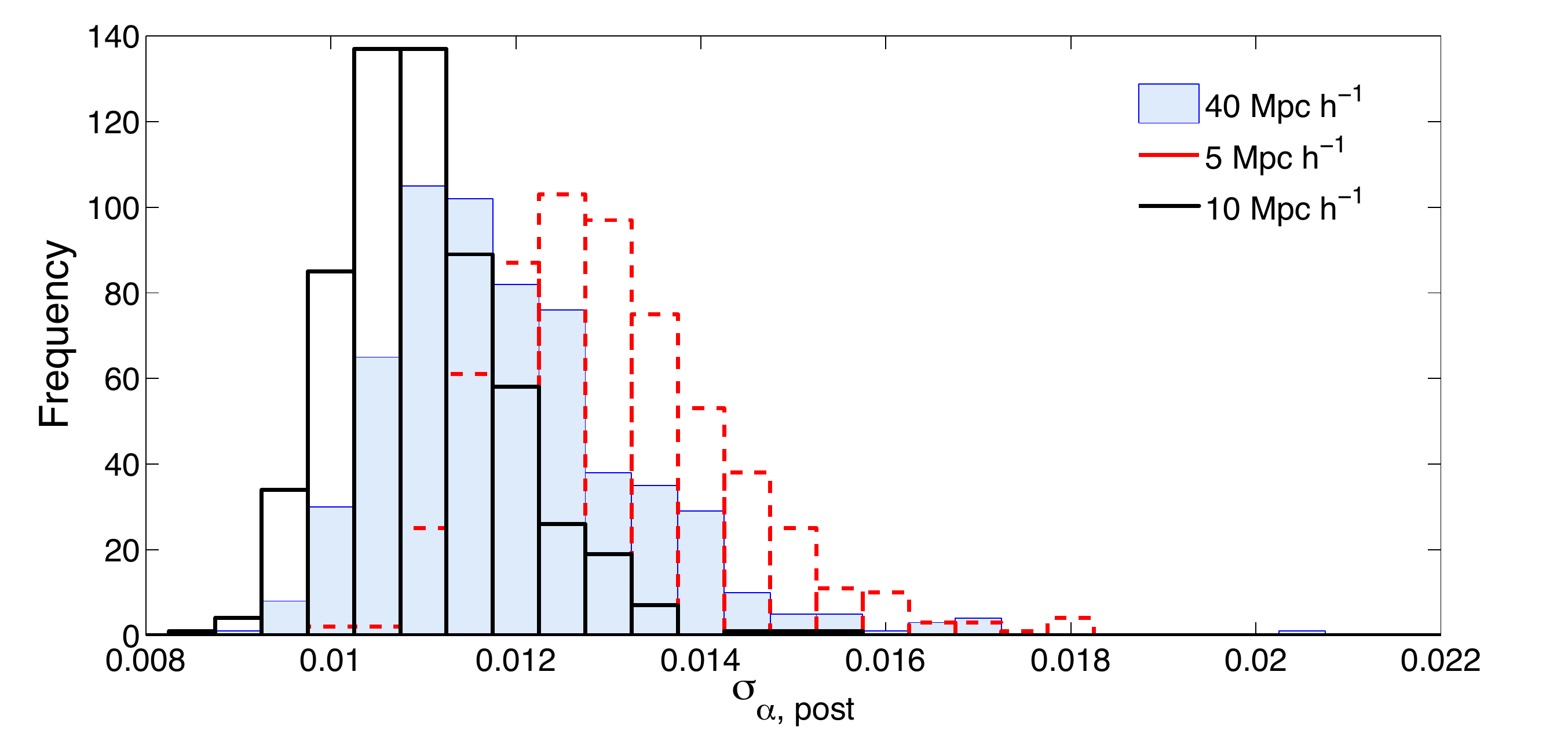}}
\endminipage\hfill
\minipage{1\textwidth}
    \centering
    \resizebox{0.46\textwidth}{!}{\includegraphics{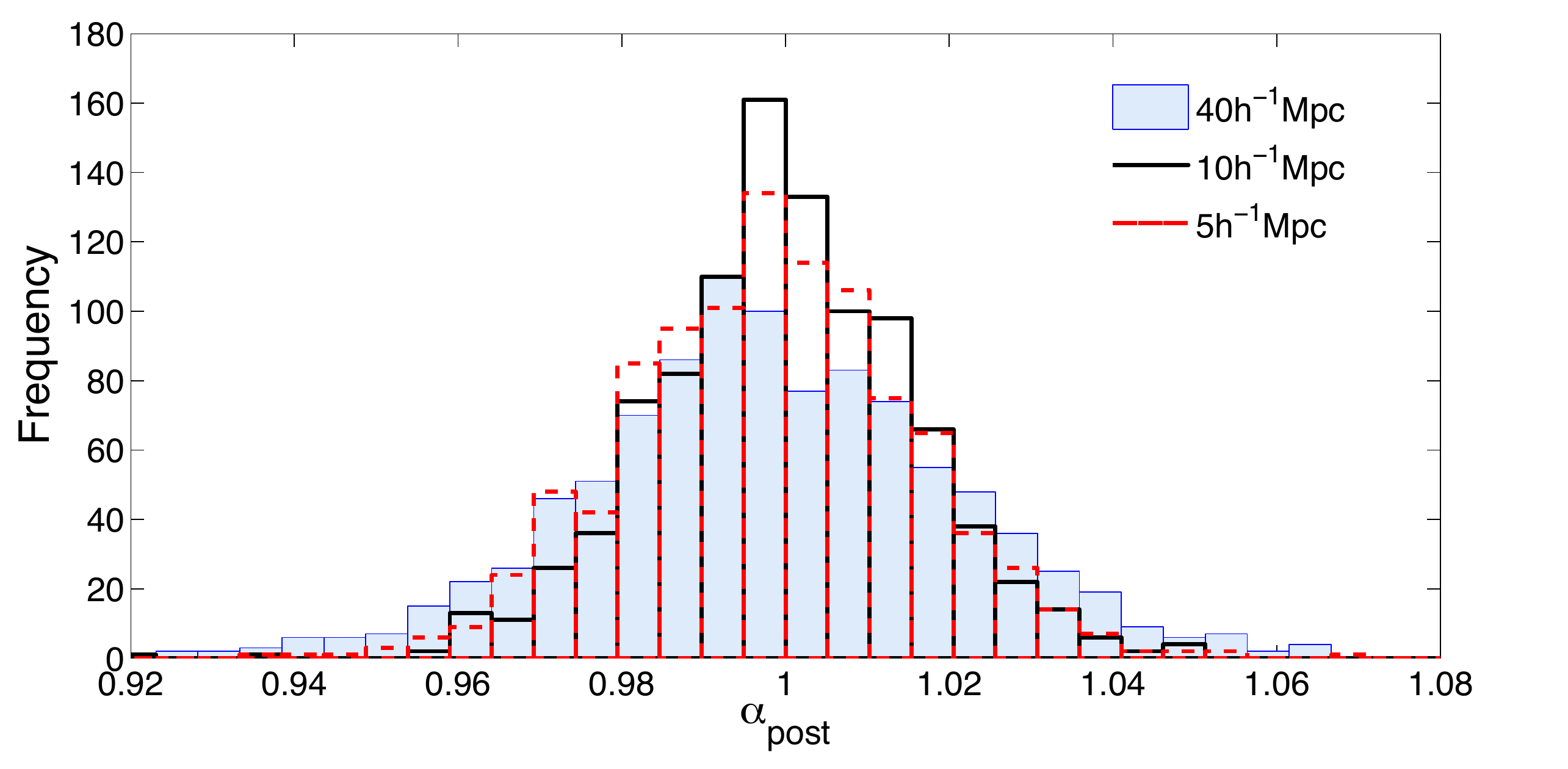}}
    \hspace{1cm}
    \resizebox{0.46\textwidth}{!}{\includegraphics{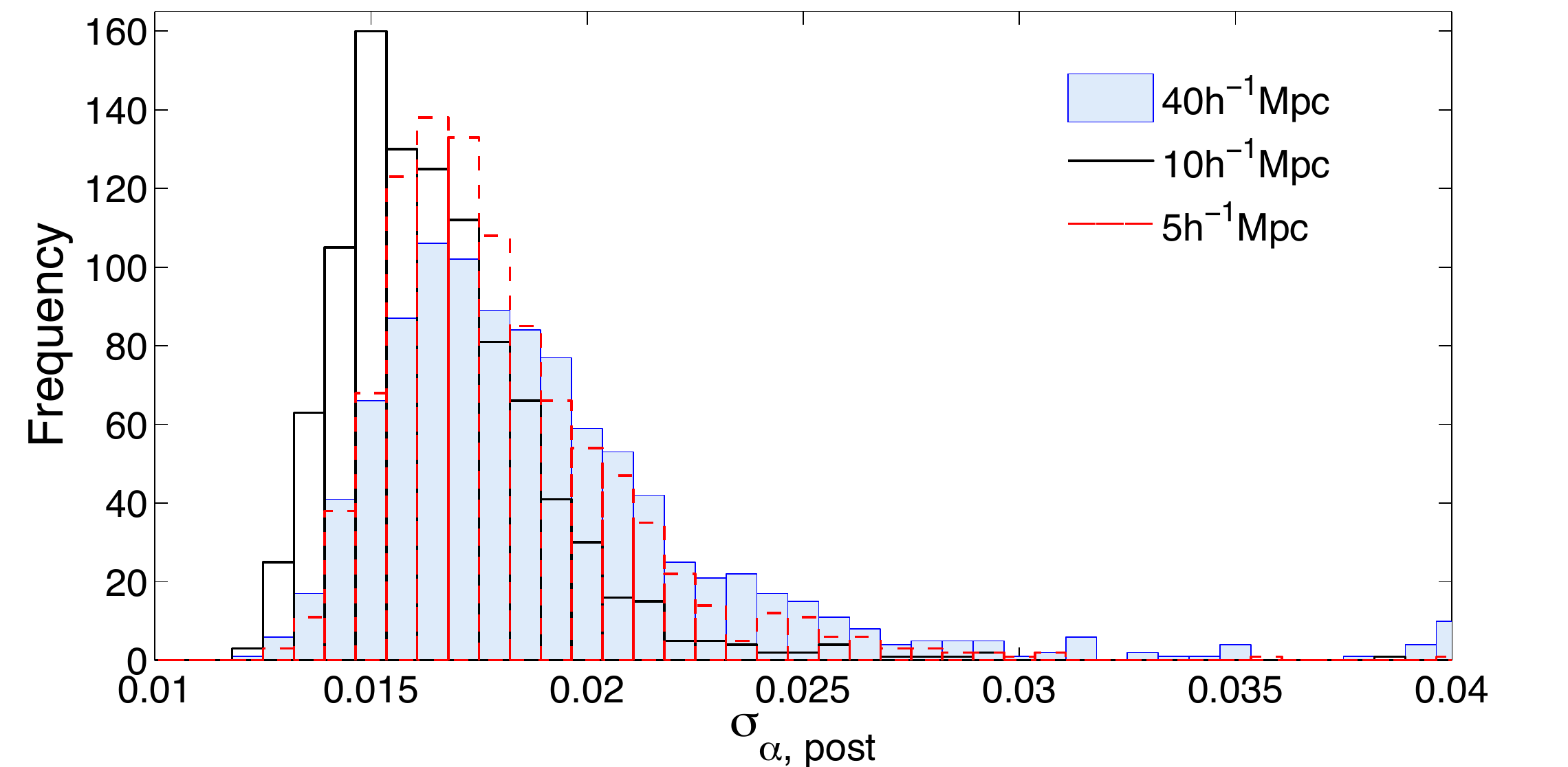}}
\endminipage\hfill
	\caption{ Distribution in $\alpha$ and $\sigma_{\alpha}$ for the CMASS (top) and LOWZ (bottom) mocks with different Gaussian smoothing kernels applied in the reconstruction.}
	\label{fig:smoothingHIST}
\end{figure*}

\clearpage


\label{lastpage}
\end{document}